\documentclass[11pt,a4paper]{article}
\pdfoutput=1
\usepackage{jheppub}
\usepackage{amsmath}
\usepackage{amssymb}
\usepackage{graphicx}
\usepackage{subcaption}
\usepackage{slashed}
\usepackage{cases}
\usepackage[toc,page]{appendix}
\usepackage{empheq}
\usepackage[normalem]{ulem} 
\usepackage{xcolor} 
\usepackage{pdfpages} 
\usepackage{tabularx}
\usepackage{multirow}

\allowdisplaybreaks

 \newcommand{\blue}{\textcolor{blue}}

\preprint{}

\begin{document}

\title{Searching for $Z'$ bosons at the P2 experiment}

\author[a]{P. S. Bhupal Dev,}
\author[b]{Werner Rodejohann,}
\author[c]{Xun-Jie Xu,}
\author[d,a]{Yongchao Zhang}
\affiliation[a]{Department of Physics and McDonnell Center for the Space Sciences,  Washington University, \\
St.\ Louis, MO 63130, USA}
\affiliation[b]{Max-Planck-Institut f\"ur Kernphysik, Saupfercheckweg 1,  69117 Heidelberg,  Germany}
\affiliation[c]{Service de Physique Th\'{e}orique, Universit\'{e} Libre de Bruxelles, Boulevard du Triomphe, CP225, 1050 Brussels, Belgium}
\affiliation[d]{School of Physics, Southeast University, Nanjing 211189, China}

\emailAdd{bdev@wustl.edu}
\emailAdd{werner.rodejohann@mpi-hd.mpg.de}
\emailAdd{xunjie.xu@ulb.ac.be}
\emailAdd{zhangyongchao@seu.edu.cn}
\date{\today}

\abstract{
The P2 experiment aims at high-precision measurements of the parity-violating asymmetry in elastic electron-proton and
electron-$^{12}$C scatterings with longitudinally polarized electrons.
We discuss here the sensitivity of P2 to new physics mediated by an additional neutral gauge boson $Z'$ of a new $U(1)'$ gauge symmetry.
If the charge assignment of the $U(1)'$ is chiral, i.e., left- and right-handed fermions have different charges under $U(1)'$,
additional parity-violation is induced directly. On the other hand, if the $U(1)'$ has a non-chiral charge assignment, additional parity-violation can be induced via mass or kinetic $Z$-$Z'$ mixing.
By comparing the P2 sensitivity to existing constraints, we show that in both cases P2 has discovery potential over a wide range of $Z'$ mass.
In particular, for chiral models, the P2 experiment can probe gauge couplings at the order of $10^{-5}$ when the $Z'$ boson is light, and  heavy $Z'$ bosons up to 79 (90) TeV in the proton  ($^{12}$C) mode. For non-chiral models with mass mixing, the P2 experiment is sensitive to mass mixing angles smaller than roughly $10^{-4}$, depending on model details and gauge coupling magnitude.

}

\maketitle

\section{ Introduction \label{sec:intro}}
Why parity is violated in elementary particle interactions remains one of the biggest mysteries in physics. Within the Standard Model (SM) of particle physics, parity-violation is caused by the weak $Z$ and $W^\pm$ bosons, which couple differently to left- and right-handed fermions. This chiral charge assignment of the SM fermions leads to various interesting phenomena and precision tests of the SM~\cite{Zyla:2020zbs}. {At the same time, given the fact that some form of beyond the SM (BSM) physics is expected on general grounds, an interesting question is whether the new BSM interactions are parity-conserving or parity-violating. Either way, BSM physics can influence parity-violating observables in reactions in which it participates. Thus, parity-violating searches provide an excellent avenue to probe BSM physics; see Ref.~\cite{Safronova:2017xyt} for a recent review.}

A classic parity-violating observable can be obtained from the scattering of polarized electrons off  unpolarized targets, yielding cross-sections $\sigma_R$ and $\sigma_L$ for right- and left-handed electrons, respectively. The parity-violating left-right asymmetry, defined as
\begin{equation}
A^{{\rm PV}} \ = \ \frac{\sigma_{R}-\sigma_{L}}{\sigma_{R}+\sigma_{L}} \, ,\label{eq:x}
\end{equation}
is then a very useful probe of parity-violation; see Ref.~\cite{Erler:2014fqa} for a review. The asymmetry $A^{\rm PV}$ has been (will be) measured with high precision in low-energy polarized electron scattering processes~\cite{Souder:2015mlu}, such as M{\o}ller scattering $e^- e^- \to e^- e^-$~\cite{Anthony:2003ub, Anthony:2005pm} (\cite{Benesch:2014bas}), as well as electron-proton~\cite{Spayde:1999qg,Aniol:2000at,Aniol:2004hp, Androic:2011rha, Androic:2018kni} (\cite{Becker:2018ggl}), electron-deuteron~\cite{Prescott:1978tm, Prescott:1979dh, Ito:2003mr, Wang:2014bba, Wang:2014guo, BalaguerRios:2016ftd}, electron-$^4$He~\cite{Aniol:2005zf}, electron-$^9$Be~\cite{Heil:1989dz},  electron-$^{12}$C~\cite{Souder:1990ia} (\cite{Becker:2018ggl})  and electron-$^{208}$Pb~\cite{Abrahamyan:2012gp} scatterings. In addition, there are precise measurements of atomic parity violation (APV) using $^{133}$Cs~\cite{Wood:1997zq, Bennett:1999pd, Porsev:2009pr, Porsev:2010de}, $^{205}$Tl~\cite{Edwards:1995zz, Vetter:1995vf}, $^{208}$Pb~\cite{Meekhof:1993zz},  $^{209}$Bi~\cite{Macpherson:1991opp} and $^{100,102,104,106}$Yb~\cite{Antypas:2018mxf}. Parity-violating asymmetries have also been measured at the high-energy colliders such as LEP, SLC, Tevatron and LHC~\cite{Zyla:2020zbs}. The interplay of these various parity-violating measurements
with new physics have been discussed, e.g.\ in Refs.\ \cite{Marciano:1990dp, Altarelli:1991ci, Erler:2003yk, Kurylov:2003zh,  Erler:2011iw,   Diener:2011jt,Davoudiasl:2012qa,Davoudiasl:2014kua,  Dev:2018sel,Long:2018fud, Carlini:2019ksi, DAmbrosio:2019tph,Arcadi:2019uif,Ghosh:2019dmi,Hong:2020dwz, Sahoo:2021thl}.



This paper deals with probing new physics using the measurements of the parity-asymmetry in elastic electron-proton or electron-$^{12}$C scatterings in the proposed P2 experiment at the upcoming Mainz Energy-recovering Superconducting Accelerator (MESA) facility~\cite{Becker:2018ggl}. The goal of P2, with start of data-taking expected in 2024, is to measure the parity-violating asymmetries for polarized electrons scattering off unpolarized protons or $^{12}$C nuclei  using a 155 MeV electron beam, where the relative uncertainties $\Delta A^{\rm PV}/A^{\rm PV}$ are expected to be 1.4\% and 0.3\% respectively~\cite{Becker:2018ggl}. Such precise measurements at low momentum transfer provide not only an important test of the SM, but also a sensitive probe of BSM physics.


One natural scenario of new physics is a  (light) $Z'$ boson that couples differently to left- and right-handed SM fermions. The
$Z'$ boson will mediate new Feynman diagrams for electron scattering off proton or nucleus (see the bottom two panels in Fig.~\ref{fig:feyn}).
%
%
%
%
%
We specifically examine how new physics in the form of a new neutral gauge boson $Z'$ can be constrained in the P2 experiment. Depending on the origin of the parity-violation in the $Z'$ couplings, the $U(1)'$ models accommodating the $Z'$ gauge boson can be classified into two categories:
\begin{itemize}
    \item {\it Chiral theories} in which left- and right-handed particles have different charge assignments under the $U(1)'$. They give a direct contribution to the parity-violating asymmetry.
    Some anomaly-free $U(1)'$ examples are given in Table~\ref{tab:chiral_models}.

    \item {\it Vector-like or non-chiral theories} in which left- and right-handed particles have identical charge assignments under the $U(1)'$. They give an indirect contribution to the parity-violating asymmetry if the $Z'$ mixes with the SM $Z$ boson. In this paper, we will first consider a generic $U(1)'$ model with either mass mixing $\sin\theta$ or kinetic mixing $\epsilon$ in the limit of the new gauge coupling $g' \to 0$, and then generalize to the $U(1)_B$ and $U(1)_{B-L}$ models (here $B$ and $L$ denote the baryon and lepton number, respectively), with three benchmark  values of $g'/\sin\theta = 0.01$, $1$ and $10$, and $\epsilon=0$.
\end{itemize}

Our results, shown in Figs.~\ref{fig:gx} to \ref{fig:U1B-L}, demonstrate that the P2 prospects of the $U(1)'$ models are rather model-dependent. However, even if all existing constraints are taken into consideration, the P2 experiment can still probe  a wide range of $Z'$ masses. For the three chiral models considered in this paper, the P2 experiment can probe gauge couplings down to $g' \sim 10^{-5}$ when the $Z'$ boson is light, as summarized in Table~\ref{tab:limits2a}. When the $Z'$ boson mass $m_{Z'}$ is large, the P2 experiment probes an effective cutoff scale $\Lambda = m_{Z'} / g'$, which for $m_{Z'}$ can go up to 79 TeV in the $e+ p$ mode, and even up to 90 TeV in the $e+^{12}$C mode (setting the $g'$ to be the perturbative limit of $4\pi$), which is well beyond direct searches at past and current  high-energy colliders.
For the non-chiral models, if there is only $Z$-$Z'$ mass mixing, in the limit of $g'Q' \to 0$, the P2 prospects in the $e+p$ mode have been precluded by APV measurements,
while P2 can probe unexplored mass mixing angles in the range of $1.1 \times 10^{-4} < \sin\theta < 0.15$ in $e+ ^{12}$C scattering, as shown in Fig.~\ref{fig:theta} and Table~\ref{tab:limits2b}.
If there is only kinetic mixing, the $Z'$ boson behaves essentially like a dark photon when it is light, and it is also severely constrained when it is heavy. Such particles  are easily accessible and often searched for. Therefore, the P2 prospects of the kinetic mixing angle $\epsilon$ have been precluded by existing limits, as presented in Fig.~\ref{fig:eps}.
For illustration, we further apply our analyses to $U(1)_B$ and $U(1)_{B-L}$ models with $g'/\sin\theta$ fixed at some benchmark values.
We find that when the gauge coupling $g'$ is sizable compared to the mass mixing angle $\sin\theta$, the P2 sensitivity to $\sin\theta$ can be significantly improved.
Our study also shows that whether $e+p$ or $e+^{12}$C scattering gives better limits depends largely on the model.

The rest of the paper is organized as follows: In Section~\ref{sec:basic} we discuss general aspects of the parity-violating asymmetry, before
discussing various chiral and non-chiral models that modify the parity-violating asymmetry in Section~\ref{sec:np}. A sensitivity study is performed in Section~\ref{sec:study}: with the procedure for obtaining the sensitivities given in Section~\ref{sec:basic}, all relevant existing limits are collected in Section~\ref{sec:limits}, and the P2 sensitivities are obtained in Section~\ref{sec:sensitivities}.
The conclusions are presented in Section~\ref{sec:concl}. More details of the axial-vector couplings of $Z'$ boson to proton and $^{12}$C are provided in Appendix~\ref{sec:gA}.

\section{ Parity-violating asymmetry in elastic electron scattering \label{sec:basic}}

\begin{figure}[t]
  \centering
  \includegraphics[height=0.35\textwidth]{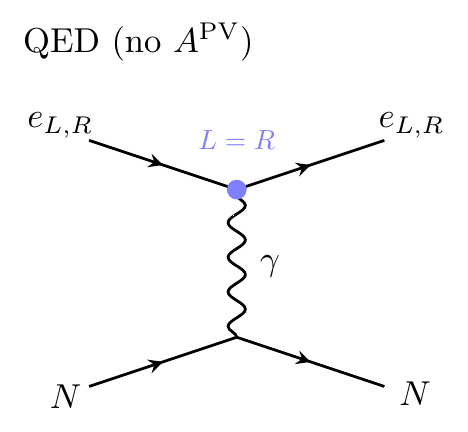}\includegraphics[height=0.35\textwidth]{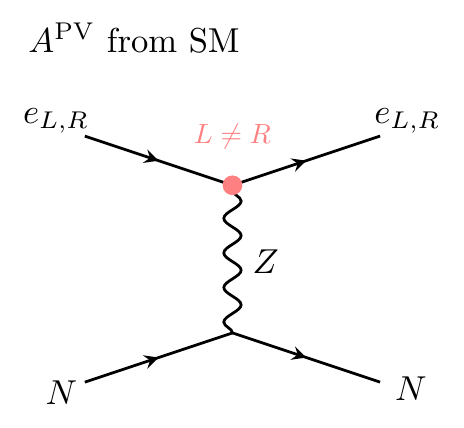}
  \vspace{5mm}
  \includegraphics[height=0.35\textwidth]{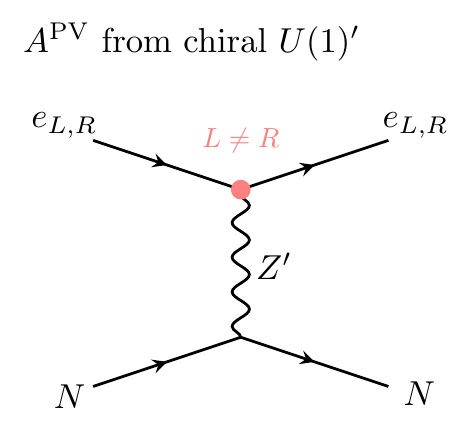}\includegraphics[height=0.35\textwidth]{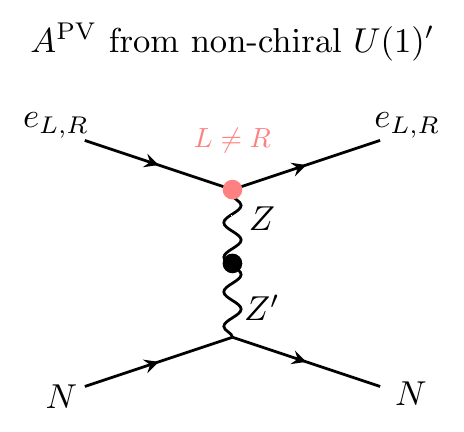}
  \caption{Leading-order processes for $A^{{\rm PV}}$ in the SM (top panels) and in the $Z'$ models considered here (bottom panels). $N$ denotes either a proton or a nucleus ($^{12}{\rm C}$ in the context of P2).
  \label{fig:feyn}}
\end{figure}

\noindent In this section, we derive the parity-violating asymmetry
$A^{{\rm PV}}$ defined in Eq.~\eqref{eq:x} for elastic electron-proton or electron-nucleus scattering.
In the SM, the leading-order contribution can be computed by evaluating
the first two diagrams in Fig.~\ref{fig:feyn}. Parity-violation enters via the $Z$ boson contribution and its interference with the parity-conserving photon diagram.

\begin{table}
    \caption{\label{tab:t}
Analytical expressions for the effective couplings of  $Z$ and $Z'$ bosons to the SM quarks and charged leptons in the SM, chiral $U(1)'$ models (from $Q_{f_L} \neq Q_{f_R}$) and non-chiral $U(1)'$ models (from the kinetic mixing $\epsilon B^{\mu\nu}F'_{\mu\nu}$ and mass mixing $\delta m^{2}Z^{\mu}Z'_{\mu}$). The last two rows are the corresponding analytic formulae of $A^{{\rm PV}}$ for electron scattering off proton or $^{12}$C. The superscript ``SM'' indicates SM contributions.
Here {$g_Z\equiv e/s_Wc_W$, $s_W\equiv \sin\theta_W$ being the weak mixing angle (and $c_W\equiv \cos\theta_W$)} and we have defined $r_{m}\equiv m_{Z'}^{2}/m_{Z}^{2}$.
See the main text for details.
}
\begin{tabular}{ccccc}
\hline\hline
models & SM & chiral models & \multicolumn{2}{c}{non-chiral models} \\ \hline
$A^{{\rm PV}}$ caused by& NC
& $Q'_{e_{L}}\neq Q'_{e_{R}}$ & $\epsilon B^{\mu\nu}F'_{\mu\nu}$
& $\delta m^{2}Z^{\mu}Z'_{\mu}$\\ \hline
$g_{e_{L}}$ & $g_{Z}\left(-\frac{1}{2}+s_{W}^{2}\right)$
& $g_{e_{L}}^{{\rm SM}}$ & $g_{e_{L}}^{{\rm SM}}+\frac{3-2r_{m}-2s_{W}^{2}}{4\left(1-r_{m}\right){}^{2}}g_{Z}s_{W}^{2}\epsilon^{2}$ & $g_{e_{L}}^{{\rm SM}}\cos\theta$\\ 

$g_{e_{R}}$ & $g_{Z}s_{W}^{2}$ & $g_{e_{R}}^{{\rm SM}}$ & $g_{e_{R}}^{{\rm SM}}+\frac{2-2r_{m}-s_{W}^{2}}{2\left(1-r_{m}\right){}^{2}}g_{Z}s_{W}^{2}\epsilon^{2}$
& $g_{e_{R}}^{{\rm SM}}\cos\theta$\\ 

$g_{u_{L}}$ & $g_{Z}\left(\frac{1}{2}-\frac{2}{3}s_{W}^{2}\right)$
& $g_{u_{L}}^{{\rm SM}}$ & $g_{u_{L}}^{{\rm SM}}-\frac{5-2r_{m}-4s_{W}^{2}}{12\left(1-r_{m}\right){}^{2}}g_{Z}s_{W}^{2}\epsilon^{2}$
& $g_{u_{L}}^{{\rm SM}}\cos\theta$\\ 

$g_{u_{R}}$ & $g_{Z}\left(-\frac{2}{3}s_{W}^{2}\right)$ & $g_{u_{R}}^{{\rm SM}}$
& $g_{u_{R}}^{{\rm SM}}-\frac{2-2r_{m}-s_{W}^{2}}{3\left(1-r_{m}\right){}^{2}}g_{Z}s_{W}^{2}\epsilon^{2}$
& $g_{u_{R}}^{{\rm SM}}\cos\theta$\\

$g_{d_{L}}$ & $g_{Z}\left(-\frac{1}{2}+\frac{1}{3}s_{W}^{2}\right)$
& $g_{d_{L}}^{{\rm SM}}$ & $g_{d_{L}}^{{\rm SM}}+\frac{1+2r_{m}-2s_{W}^{2}}{12\left(1-r_{m}\right){}^{2}}g_{Z}s_{W}^{2}\epsilon^{2}$ & $g_{d_{L}}^{{\rm SM}}\cos\theta$\\ 

$g_{d_{R}}$ & $g_{Z}\left(\frac{1}{3}s_{W}^{2}\right)$ & $g_{d_{R}}^{{\rm SM}}$
& $g_{d_{R}}^{{\rm SM}}+\frac{2-2r_{m}-s_{W}^{2}}{6\left(1-r_{m}\right){}^{2}}g_{Z}s_{W}^{2}\epsilon^{2}$
& $g_{d_{R}}^{{\rm SM}}\cos\theta$\\ \hline 
$g'_{e_{L}}$ & 0 & $Q'_{e_{L}}g'$ & $\frac{c_{W}^{2}-r_{m}/2}{1-r_{m}}\epsilon g_{Z}s_{W}$
& $g_{e_{L}}^{{\rm SM}}\sin\theta$\\

$g'_{e_{R}}$ & 0 & $Q'_{e_{R}}g'$ & $\frac{c_{W}^{2}-r_{m}}{1-r_{m}}\epsilon g_{Z}s_{W}$
& $g_{e_{R}}^{{\rm SM}}\sin\theta$\\

$g'_{u_{L}}$ & 0 & $Q'_{u_{L}}g'$ & $\frac{r_{m}-4c_{W}^{2}}{6\left(1-r_{m}\right)}\epsilon g_{Z}s_{W}$
& $g_{u_{L}}^{{\rm SM}}\sin\theta$\\

$g'_{u_{R}}$ & 0 & $Q'_{u_{R}}g'$ & $\frac{2(r_{m}-c_{W}^{2})}{3\left(1-r_{m}\right)}\epsilon g_{Z}s_{W}$
& $g_{u_{R}}^{{\rm SM}}\sin\theta$\\

$g'_{d_{L}}$ & 0 & $Q'_{d_{L}}g'$ & $\frac{r_{m}+2c_{W}^{2}}{6\left(1-r_{m}\right)}\epsilon g_{Z}s_{W}$&
$g_{d_{L}}^{{\rm SM}}\sin\theta$\\

$g'_{d_{R}}$ & 0 & $Q'_{d_{R}}g'$ & $\frac{c_{W}^{2}-r_{m}}{3\left(1-r_{m}\right)}\epsilon g_{Z}s_{W}$&
$g_{d_{R}}^{{\rm SM}}\sin\theta$\\ 
\hline
$A^{{\rm PV}}$ ($p$) & Eq.~\eqref{eq:m-2} & Eq.~\eqref{eq:m-12}
& Eq.~\eqref{eq:APV_eps} & Eq.~\eqref{eq:m-5}\\ 
$A^{{\rm PV}}$ ($^{12}{\rm C}$) & Eq.~\eqref{eq:m-2-1} & Eq.~\eqref{eq:m-12}
 & Eq.~\eqref{eq:APV_epsC12} & Eq.~\eqref{eq:m-5}  \\ 
\hline \hline
\end{tabular}
\end{table}

A new neutral gauge
boson $Z'$ could also contribute to $A^{{\rm PV}}$ either directly
or indirectly via mixing with the SM $Z$ boson, depending on whether
the $Z'$ couplings to $e_{L}$ and $e_{R}$ are different or not,
respectively. This is illustrated by the diagrams in the bottom two panels of Fig.~\ref{fig:feyn}.
In the mixing case,
we need to canonicalize kinetic terms or diagonalize mass terms so
that both $Z$ and $Z'$ are mass eigenstates, leading to effective
parity-violating couplings of the $Z'$.
Besides, in the
presence of $Z$-$Z'$ mixing, the $Z$ couplings may also deviate
from the SM values. This is discussed in detail in Sec.\ \ref{sec:np}. In general, we can consider the following Lagrangian, which contains the neutral current (NC) interactions of the SM $Z$ boson and the most general interactions of a $Z'$ with polarized electrons and a target nucleus $N$:
\begin{eqnarray}\label{eq:L}
{\cal L} & \ \supset \ & Z_{\mu}(g_{e_{L}}\overline{e_{L}}\gamma^{\mu}e_{L}+g_{e_{R}}\overline{e_{R}}\gamma^{\mu}e_{R})+Z{}_{\mu}(g_{V}\overline{N}\gamma^{\mu}N+g_{A}\overline{N}\gamma^{\mu}\gamma^{5}N)\nonumber \\
 & & \quad  +  Z'_{\mu}(g'_{e_{L}}\overline{e_{L}}\gamma^{\mu}e_{L}+g'_{e_{R}}\overline{e_{R}}\gamma^{\mu}e_{R})+Z'_{\mu}(g'_{V}\overline{N}\gamma^{\mu}N+g'_{A}\overline{N}\gamma^{\mu}\gamma^{5}N). \label{eq:x-1}
\end{eqnarray}
Here $Z$ and $Z'$ are  mass eigenstates with  masses denoted
by $m_{Z}$ and $m_{Z'}$, $g_{e_{L}}$ and $g_{e_{R}}$ ($g'_{e_{L}}$
and $g'_{e_{R}}$) are the effective couplings of $Z$ ($Z'$) to
$e_{L}$ and $e_{R}$, $g_{V}$ and $g_{A}$ ($g'_{V}$ and $g'_{A}$)
are vector and axial-vector couplings of $Z$ ($Z'$) to a nucleus, respectively. The vector couplings can be obtained by simply adding up the fundamental couplings to quarks in a nucleus with ${\cal N}$ neutrons and ${\cal Z}$ protons as
follows:
\begin{eqnarray}
g_{V} & \ = \ & {\cal Z}\left(g_{u_{L}}+g_{u_{R}}+\frac{1}{2}g_{d_{L}}+\frac{1}{2}g_{d_{R}}\right)+{\cal N}\left(g_{d_{L}}+g_{d_{R}}+\frac{1}{2}g_{u_{L}}+\frac{1}{2}g_{u_{R}}\right),\label{eq:x-2}\\
g'_{V} & \ = \ & {\cal Z}\left(g'_{u_{L}}+g'_{u_{R}}+\frac{1}{2}g'_{d_{L}}+\frac{1}{2}g'_{d_{R}}\right)+{\cal N}\left(g'_{d_{L}}+g'_{d_{R}}+\frac{1}{2}g'_{u_{L}}+\frac{1}{2}g'_{u_{R}}\right). \label{eq:x-10}
\end{eqnarray}
Here the couplings to chiral quarks ($g_{u_{L}}$, $g_{u_{R}}$,
$g'_{d_{R}}$, etc.) are defined in a way similar to $g_{e_{L}}$
and $g_{e_{R}}$ in Eq.~\eqref{eq:x-1}. In Tab.~\ref{tab:t} we
list the SM values of these couplings as well as new physics values
which will be derived later in Sec.\ \ref{sec:np}.

As for the axial-vector couplings ($g_{A}$ and $g'_{A}$), albeit
not calculable from first principles, their contributions to $A^{{\rm PV}}$
are suppressed by electron energy over target mass, as we show in Appendix~\ref{sec:gA}.
If we therefore ignore the contribution of $g_{A}$ and $g'_{A}$,
the amplitude
of $e_{L,R}^{-}+N\rightarrow e_{L,R}^{-}+N$ reads
\begin{eqnarray}
i{\cal M}_{L,R} & \ \propto \ & [\overline{u_{3}}\gamma^{\mu}P_{L,R}u_{1}]\left[\overline{u_{4}}\gamma_{\mu}G_{L,R}u_{2}\right] \, ,\label{eq:x-7}
\end{eqnarray}
where $\overline{u_{3}}$ ($u_{1}$) and $\overline{u_{4}}$ ($u_{2}$)
denote the final (initial) electron and nucleon states, and
\begin{equation}
G_{L,R} \ = \ \frac{-e^{2}}{q^{2}}+\frac{g_{e_{L},e_{R}}g_{V}}{q^{2}-m_{Z}^{2}}+\frac{g'_{e_{L},e_{R}}g'_{V}}{q^{2}-m_{Z'}^{2}}\label{eq:x-6}
\end{equation}
contains contributions from the $t$-channel $\gamma$, $Z$, and
$Z'$ diagrams, respectively. Applying the standard trace technology,
it is straightforward to obtain
\begin{equation}
\frac{|{\cal M}_{L}|^{2}}{|{\cal M}_{R}|^{2}} \ = \ \frac{G_{L}^{2}}{G_{R}^{2}} \, ,\label{eq:x-5}
\end{equation}
which, according to Eq.~(\ref{eq:x}), implies
\begin{equation}
A^{{\rm PV}} \  = \ \frac{G_{R}^{2}-G_{L}^{2}}{G_{R}^{2}+G_{L}^{2}} \, .\label{eq:m}
\end{equation}
Now substituting Eq.~(\ref{eq:x-6}) in Eq.~(\ref{eq:m}), we obtain
\begin{equation}
A^{{\rm PV}} \ \approx \ \frac{\left(g_{e_{L}}-g_{e_{R}}\right)g_{V}}{4\pi\alpha}\frac{Q^{2}}{m_{Z}^{2}}+\frac{\left(g'_{e_{L}}-g'_{e_{R}}\right)g'_{V}}{4\pi\alpha}\frac{1}{1+m_{Z'}^{2}/Q^{2}} \, ,\label{eq:m-1}
\end{equation}
where $Q^{2}\equiv-q^{2}>0$ and $\alpha\equiv e^2/4\pi$ is the fine-structure constant. In deriving  Eq.~\eqref{eq:m-1}, we have made
 the following approximations:
\begin{equation}
\frac{Q^{2}}{m_{Z}^{2}} \ \ll \ 1 \, , \qquad {\rm and} \qquad  \frac{g'_{e}g'_{V}}{1+m_{Z'}^{2}/Q^{2}} \ \ll \ 1 \, ,\label{eq:m-6}
\end{equation}
where $g'_{e}\sim\max(g'_{e_{L}},\ g'_{e_{R}})$. The exact expression of $A^{{\rm PV}}$ is lengthy and not very helpful. Within a theory containing an additional $Z'$ boson, once the six effective couplings ($g_{e_{L}}$, $g_{e_{R}}$, $g_{V}$, $g'_{e_{L}}$, $g'_{e_{R}}$,
$g'_{V}$) are known,  $A^{{\rm PV}}$ can be obtained using Eq.~\eqref{eq:m-1} and confronted with its existing or future constraints, as exemplified in the following two sections.

We stress that Eq.~\eqref{eq:m-1} is obtained from the most general Lagrangian in Eq.\ (\ref{eq:L}). In the limit of $g'_{e_L} = g'_{e_R}$, $g_V'\to 0$ or $m_{Z'}\to \infty$, the BSM contributions are vanishing, and taking the SM values in Tab.~\ref{tab:t} and using Eq.~\eqref{eq:x-2},
we can easily reproduce the leading-order SM expressions of the parity-violating asymmetries  $A^{{\rm PV}}$~\cite{Becker:2018ggl}:
\begin{eqnarray}
A_{{\rm SM}}^{{\rm PV}}(e+p) & \ \approx \ & -\frac{G_{F}Q^{2}\left(1-4s_{W}^{2}\right)}{4\sqrt{2}\pi\alpha} \, ,\label{eq:m-2}\\
A_{{\rm SM}}^{{\rm PV}}(e+{}^{12}{\rm C}) & \ \approx \ & \frac{3\sqrt{2}G_{F}Q^{2}s_{W}^{2}}{\pi\alpha} \, ,\label{eq:m-2-1}
\end{eqnarray}
for electron-proton scattering (${\cal N}=0$ and ${\cal Z}=1$) and
electron-$^{12}$C scattering (${\cal N}={\cal Z}=6$), respectively. Here $G_F=\pi \alpha/(\sqrt 2 m_W^2 s_W^2)$ is the Fermi constant.




\section{New physics effects}\label{sec:np}

This section deals with obtaining the couplings of an additional $Z'$ boson with the SM fermions in some representative $U(1)'$ models to study the P2 prospects.
We first discuss three chiral $U(1)'$ models with explicit parity-violation, and then turn to parity-conserving $U(1)'$ models with mass or kinetic $Z$-$Z'$ mixing.

\subsection{Chiral $U(1)'$ models}

First let us consider new $U(1)'$ gauge symmetries under
which $e_{L}$ and $e_{R}$ have different charges. We refer to such
models as chiral $U(1)'$ models. Anomaly cancellation usually requires the existence of three right-handed neutrinos $\nu_R$ (for simplicity we have neglected the generation indices $i = 1,\, 2,\, 3$), and implies the following charge assignments of chiral fermions~\cite{Oda:2015gna,Campos:2017dgc}:
\begin{align}
Q_{u_{R}}'   & \ = \ x, &
Q_{d_{R}}'& \ = \ y, &
Q_{\nu_{R}}' &\ = \ -x-2y, \nonumber \\
Q_{e_{R}}' & \ = \ -2x-y , &
Q'_{Q_{L}} & \ = \  \frac{1}{2} (x+y), &
Q_{L_{L}}'& \ = \  -\frac{3}{2} (x+y) \,.
\label{eq:m-8}
\end{align}
Here $Q_L = (u_L, \; d_L)^{\sf T}$ and $L_L = (\nu_L, \; e_L)^{\sf T}$ are respectively the left-handed quark and lepton doublets,  and $Q_{f}'$ is the $U(1)'$ charge of the chiral fermion $f$.
The charges  $x$ and $y$ can be arbitrary integers or fractional numbers.
However, from the model building perspective, simple integers or fractions are preferred in order to construct simple (low-dimensional) gauge-invariant operators.
For simplicity, we will assume that the right-handed neutrinos are heavier than $m_{Z'}/2$, such that the decay channel $Z' \to \nu_R \bar\nu_R$ is kinematically forbidden and  the $\nu_R$ do not have any effect on the P2 sensitivities.

The simplest example is perhaps the $U(1)'_{R}$ model given in Tab.~\ref{tab:chiral_models}. In
this model, all right-handed fermions are charged under $U(1)'_{R}$,
and none of the left-handed fermions are, i.e. $(x,y)=(1,-1)$ in Eq.~\eqref{eq:m-8}; see e.g.~Refs.~\cite{Alikhanov:2019drg,Dutta:2019fxn, Jana:2019mez}.   One might also consider the opposite case where
all left-handed fermions are charged and none of the  right-handed fermions
are. Unfortunately for such assignment it is impossible to achieve
chiral anomaly cancellation. If, for instance, $e_{R}$ is not charged under the
chiral $U(1)'$, which corresponds to
$x/y=-1/2$ in Eq.~\eqref{eq:m-8}, then the only solution is where the other right-handed fermions are charged for
anomaly cancellation. This is the $U(1)'_{L}$ case shown in Tab.~\ref{tab:chiral_models}.
In addition to $U(1)'_{R}$ and $U(1)'_{L}$, we choose another solution with $(x,y)=(0,1)$, where right-handed up-quarks do not couple of $Z'$, and refer to it as $U(1)_{X}'$ in Tab.~\ref{tab:chiral_models}.
Obviously there are many more possibilities. Our choice of the three models summarized
in Tab.~\ref{tab:chiral_models} is motivated by the fact that they are typical for different corners of the parameter space to be explored at P2. For instance, the $U(1)_R$ model will not contribute to the asymmetry in electron-$^{12}$C scattering (cf.\ Eq.~(\ref{eq:m-11}) below).

\begin{table}
    \centering
    \caption{Quantum numbers of fermions in the three chiral $U(1)'$ examples considered here. 
    \label{tab:chiral_models}}
\begin{tabular}{cc|cccc}
\hline \hline
\multirow{3}{*}{model}  & \multirow{3}{*}{$(x,y)$} & \multicolumn{4}{c}{$U(1)'$ charge assignment} \\ \cline{3-6}
& & $(\nu_{L},e_{L})$ & $(\nu_{R},e_{R})$ & $(u_{L},d_{L})$ & $(u_{R},d_{R})$ \\
\cline{3-6}
 & & $-\frac{3}{2} (x+y)$ & $-(x+2y,\ 2x+y)$ & $\frac{1}{2} (x+y)$ &
$(x,\ y)$ \\[2mm] \hline
$U(1)_{L}'$ & $(-\frac{2}{3},\frac{4}{3})$ & $-1$ & $(-2,0)$ & $\frac{1}{3}$ & $(-\frac{2}{3},\frac{4}{3})$
\\[1mm]
$U(1)_{R}'$ & $(1,-1)$ &  $0$ & $(1,-1)$ & $0$ & $(1,-1)$ \\[1mm]
$U(1)_{X}'$ & $(0,1)$ &  $-\frac{3}{2}$ & $(-2,-1)$ & $\frac{1}{2}$ &
$(0,1)$ \\[1mm]
\hline\hline
\end{tabular}
\end{table}

Computing $A^{{\rm PV}}$ for chiral $U(1)'$ models is now straightforward.
In the absence of $Z$-$Z'$ mixing, the $Z$ couplings are not modified and the $Z'$ couplings in Eq.~\eqref{eq:x-1}
are given by
\begin{equation}
g'_{f} \ = \ g'Q'_{f} \, ,\label{eq:m-9}
\end{equation}
where $g'$ is the fundamental gauge coupling of $U(1)'$.
Using Eq.~\eqref{eq:m-1}, we obtain
\begin{equation}
\Delta A^{{\rm PV}} \ \equiv \ A^{{\rm PV}}-A_{{\rm SM}}^{{\rm PV}} \ \approx \ \frac{g'^{2}Q'_{N}Q'_{LR}}{4\pi\alpha}\frac{1}{1+m_{Z'}^{2}/Q^{2}} \, ,\label{eq:m-12}
\end{equation}
where $Q'_{LR}\equiv Q'_{e_{L}}-Q'_{e_{R}}$, and $Q'_{N}$ is either the weak charge $Q'_p$ of the proton or the weak charge $Q'_{^{12}{\text{C}}} $ of $^{12}$C, with
\begin{align}
Q'_{p} & \ \equiv \  Q'_{u_{L}}+Q'_{u_{R}}+\frac{1}{2}\left(Q'_{d_{L}}+Q'_{d_{R}}\right) \, ,\label{eq:m-10}\\
Q'_{^{12}{\rm C}} & \ \equiv \  9\left(Q'_{u_{L}}+Q'_{u_{R}}+Q'_{d_{L}}+Q'_{d_{R}}\right) \, .
\label{eq:m-11}
\end{align}
From Tab.~\ref{tab:chiral_models}, we note that in the $U(1)'_{R}$ model, $Q'_{^{12}{\rm C}}$ vanishes, which implies
that electron-$^{12}$C scattering has no sensitivity to $Z'$ in this model.

Here we comment on the potential effect of incoherence in $e + {}^{12}{\rm C}$
scattering and the validity of summing up the charges in Eq.~\eqref{eq:m-11}.
In the P2 experiment, $Q \approx 93 \ {\rm MeV}$ is comparable
to the inverse of the $^{12}{\rm C}$ nucleus radius and there is
a considerable amount of loss of coherence, which is usually taken
into account by including a form factor   depending on the nucleon
distributions. Due to the small difference between neutron and proton
distributions, two types of form factors are often considered, namely
weak and charge form factors, denoted respectively by $F_{W}(Q^{2})$ and $F_{{\rm Ch}}(Q^{2})$.
The form factor $F_{\rm Ch} (Q^2)$ depends on the proton distribution in the nucleus, while $F_{W}(Q^2)$ is determined mainly by the neutron distribution,  as the coupling of $Z$ to neutrons is significantly larger than the coupling of $Z$ to protons. In presence of the couplings of $Z'$ to protons and
neutrons, the cross sections $\sigma_{L,\,R}$ depend on the combination of
$F_{W} (Q^2) - F_{{\rm Ch}} (Q^2)$. Using the Helm analytic approximation
for form factors~\cite{Helm:1956zz}, we find that the two form factors $F_{W,\, {\rm Ch}}  (Q^2)$ are both around $0.75$ with percent-level uncertainties for $^{12}{\rm C}$ at $Q \approx 93$ MeV. This implies that with the approximation of $F_{W} (Q^2) \simeq F_{\rm Ch} (Q^2)$ the effect of incoherence cancels out for $A^{\rm PV}$, which justifies Eq.~\eqref{eq:m-11}. It should be noted that this justification does not depend on the details of $U(1)'$ models and applies also to all the non-chiral $U(1)'$ models discussed below.

\subsection{Non-chiral $U(1)'$ models and mixing-induced couplings}

\noindent There are plenty of non-chiral $U(1)'$ models with parity-conserving
charge assignments. For instance, the $U(1)_{B-L}$ model~\cite{Davidson:1978pm, Mohapatra:1980qe, Buchmuller:1991ce}, which assigns
all leptons (both left- and right-handed) a charge of $-1$ and all
quarks a charge of $\frac{1}{3}$, is among the most extensively studied
models in the literature. Although the non-chiral charge assignments
do not cause any additional source of asymmetry in polarized electron scattering,  constraints  on non-chiral $U(1)'$ models are possible in the presence of $Z$-$Z'$ mixing, as illustrated by the lower right diagram in Fig.~\ref{fig:feyn}.

Let us consider that the $Z$-$Z'$ mixing arises from the following terms:
\begin{equation}
{\cal L} \ \supset \ -\frac{\epsilon}{2}B^{\mu\nu}F'_{\mu\nu}+\delta m^{2}\hat{Z}^{\mu}\hat{Z}'_{\mu} \, ,\label{eq:x-3}
\end{equation}
where $B^{\mu\nu}$ and $F'_{\mu\nu}$ are the  field strength tensors
of hypercharge $U(1)_{Y}$ and $U(1)'$, respectively. The second term is a mass-mixing term
which  exists if the symmetry breaking of $U(1)'$ is not
decoupled from the electroweak symmetry breaking. We define a mass
mixing parameter $\theta$ by
\begin{equation}
\tan\theta \ \equiv \ \frac{\delta m^{2}}{m_{Z'}^{2}-m_{Z}^{2}} \, ,
\label{eq:m-3-2}
\end{equation}
where $m_{Z}^{2}$ is the $Z$ boson mass. Note that in the presence of Eq.~(\ref{eq:x-3}), the kinetic terms are not canonical and the mass matrix is not diagonal.
One needs to perform necessary transformations and redefine $Z$
and $Z'$ as the physical mass eigenstates. To avoid potential
confusion, we have denoted the original states as $\hat{Z}$ and
$\hat{Z}'$ in Eq.~(\ref{eq:x-3}). Following the notation in Ref.~\cite{Lindner:2018kjo}, the couplings in the  physical basis can be written as
\begin{align}
{\cal L} & \ \supset \ Z_{\mu}\left[J_{{\rm NC}}^{\mu}\cos\beta-J_{X}^{\mu}\sin\beta\right] 
+Z'_{\mu}\left[J_{{\rm NC}}^{\mu}\sin\beta+J_{X}^{\mu}\cos\beta\right] \, ,\label{eq:m-13}
\end{align}
where the effective $Z$-$Z'$ mixing angle is
\begin{eqnarray}
\tan\beta  \ \approx \ \tan\theta+\frac{\epsilon s_{W}}{r_m-1}+{\cal O}(\theta^{2},\ \epsilon^{2}) \, ,
\label{eq:m-14}
\end{eqnarray}
where we have defined the mass ratio
\begin{eqnarray}
r_{m} \ \equiv \ \frac{m_{Z'}^{2}}{m_{Z}^{2}} \, .
\end{eqnarray}
Eq.~(\ref{eq:m-14}) incorporates contributions from the mass mixing $\theta$ and the kinetic mixing $\epsilon$.
The SM  NC $J_{{\rm NC}}^{\mu}$ and the $Z'$-induced neutral current $J_{X}^{\mu}$ in Eq.~(\ref{eq:m-14}) are defined respectively as
\begin{align}
J_{{\rm NC}}^{\mu} & \ = \ \sum_{f}g_{f}^{{\rm SM}}\overline{f}\gamma^{\mu}f \,, \\
J_{X}^{\mu} & \  = \ \sum_{f}\frac{\overline{f}\gamma^{\mu}f}{\sqrt{1-\epsilon^{2}}}\left[g'Q_{f}'-g_{Z}\epsilon s_{W}Y_{f}\right] \, ,
\label{eq:m-15}
\end{align}
with $g_{f}^{{\rm SM}}$ the SM gauge couplings for the chiral fermions $f$ collected in the second column of Tab.~\ref{tab:t}. In  Eq.~\eqref{eq:m-15}, $g_{Z}\equiv {g}/{c_{W}}= {\sqrt{4\pi\alpha}}/{s_{W}c_{W}}$ is a SM coupling, and $Y_f$ is the SM $U(1)_{Y}$ charge for the fermions.


From Eqs.~\eqref{eq:m-13}-\eqref{eq:m-15} it is straightforward
to derive the effective couplings of $Z$ and $Z'$ to fermions; these are  summarized  in Tab.~\ref{tab:t}. Note that for $Z$ couplings
we have expanded the results to  order  $\epsilon^{2}$ because
the $Z$ diagram with one vertex modified by $\epsilon^{2}$ and the
$Z'$ diagram with two vertices proportional to $\epsilon$ have comparable
contributions to the parity-violating asymmetry.

Now using the effective couplings and Eq.~(\ref{eq:m-1}), we can
compute the mixing-induced contribution to the parity-violating asymmetry, $\Delta A^{{\rm PV}}\equiv A^{{\rm PV}}-A^{{\rm PV}}_{\rm SM}$, for non-chiral $U(1)'$
models. The result, in the presence of only mass-mixing, is
\begin{equation}
\frac{\Delta A^{{\rm PV}}}{A_{{\rm SM}}^{{\rm PV}}} \ = \ \sin^2{\theta}\left(\frac{m_{Z}^{2}}{m_{Z'}^{2}+Q^{2}}-1\right).
\label{eq:m-5}
\end{equation}
For only kinetic mixing, the result is
\begin{eqnarray}
\label{eq:APV_eps}
\frac{\Delta A^{{\rm PV}}}{A_{{\rm SM}}^{{\rm PV}}}(e+p) & \ = \ & \frac{m_{Z'}^{2}(1-r_{m})\left(1-4s_{W}^{2}\right)+Q^{2}\left(4r_{m}s_{W}^{2}+2r_{m}-3\right)}{(1-r_{m})^{2}\left(1-4s_{W}^{2}\right)\left(m_{Z'}^{2}+Q^{2}\right)}s_{W}^{2}\epsilon^{2} \,, \\
\frac{\Delta A^{{\rm PV}}}{A_{{\rm SM}}^{{\rm PV}}}(e+{}^{12}{\rm C}) & \ = \ & \frac{m_{Z'}^{2}s_{W}^{2}(1-r_{m})+Q^{2}\left(1-r_{m}-r_{m}s_{W}^{2}\right)}{(1-r_{m})^{2}\left(m_{Z'}^{2}+Q^{2}\right)}\epsilon^{2} \, ,
\label{eq:APV_epsC12}
\end{eqnarray}
for electron-proton and
electron-$^{12}$C scattering, respectively.

When both mass and kinetic mixing are present, one can add their contributions
linearly as long as both $\epsilon$ and $\theta$ are perturbatively
small. In the above results, we have neglected the contribution of
the fundamental gauge coupling $g'$ in Eq.~\eqref{eq:m-15}. Although
for non-chiral $U(1)'$ models this part does not account for parity
violation, sizable values of $g'$ do affect the P2 sensitivity by
changing the effective couplings to nucleons if $g'$ is comparable
to $\theta$ or $\epsilon$. This effect can be readily included by
adding a correction of $g'Q_{f}'$ to the mixing-induced couplings
as long as both are perturbatively small. In this case, the P2 sensitivity
to $\epsilon$ and $\theta$ becomes model-dependent. We select two
non-chiral $U(1)'$ models when the model-dependent details are required,
$U(1)_{B-L}$ as aforementioned, and $U(1)_{B}$ which gauges the
baryon number only \cite{Foot:1989ts}.
For these models with sizable $g'$ and $\theta$, the new physics
contributions can be formulated as
\begin{equation}
\frac{\Delta A^{{\rm PV}}}{A_{{\rm SM}}^{{\rm PV}}} \ = \  \sin^2\theta\left[\frac{m_{Z}^{2} (1+R) }{m_{Z'}^{2}+Q^{2}}-1\right],
\label{eq:n-1}
\end{equation}
where $R$ takes the following expressions for the $U(1)_{B}$ model:
\begin{eqnarray}
\label{eqn:Rep}
R_{e+p} [U(1)_B] & \ = \ & \frac{12g'}{g_{Z}  \sin\theta(1-4s_{W}^{2})} \,, \\
\label{eqn:Rec}
R_{e+^{12}\text{C}} [U(1)_B] & \ = \ & -\frac{6g'}{g_{Z} \sin\theta s_{W}^{2}} \,.
\end{eqnarray}
For $U(1)_{B-L}$, we simply need to multiply the $R$ factors in Eqs.~(\ref{eqn:Rep}) and (\ref{eqn:Rec}) by a factor of $-1/3$, i.e.
\begin{eqnarray}
\label{eqn:Rfactor}
R_{} [U(1)_{B-L}] \ = \ -\frac13
R_{} [U(1)_{B}] \,.
\end{eqnarray}

\section{P2 sensitivity study}\label{sec:study}
In this section, we derive the P2 sensitivity for the $U(1)$ models discussed in the previous section.
\subsection{Method}
\label{sec:method}

\noindent The P2 experiment will measure polarized electron scattering
with an electron beam of 155 MeV \cite{Kempf:2020bfx}. The momentum transfer $Q^{2}$ in this process can be determined by
\begin{equation}\label{eq:Q2}
Q^{2} \ \approx \ 4E_{i}E_{f}\sin(\theta_{f}/2) \, ,
\end{equation}
where $\theta_{f}$ is the scattering angle of the outgoing electron
with respect to the incoming one, and $E_{f}\approx E_{i}\approx155$
MeV are the final/initial electron energies. The angular acceptance
of the detector is
\begin{equation}
\theta_{f} \ \in \ \left[ \overline{\theta}_{f}-\frac{1}{2}\delta\theta_{f},\ \overline{\theta}_{f}+\frac{1}{2}\delta\theta_{f} \right], \quad \text{with} \quad  \overline{\theta}_{f}=35^{\circ},\ \delta\theta_{f}=20^{\circ} \,.
\label{eq:m-17}
\end{equation}
In our analysis we take $Q^{2}\approx(93\ {\rm MeV})^{2}$
evaluated from the central value $\overline{\theta}_{f}$.
For $e+p$ scattering, the expected value of $A^{{\rm PV}}$ is
$39.94\times10^{-9}$ and the P2 experiment will be able to measure
it with an uncertainty of $\Delta A^{{\rm PV}}=0.56\times10^{-9}$,
which corresponds to a relative uncertainty of
\begin{equation}
\frac{\Delta A^{{\rm PV}}}{A^{{\rm PV}}} \ \approx \ 1.4\% \, ,\ \
(e+p \text{ scattering})
.\label{eq:m-18}
\end{equation}
This would imply a relative uncertainty of $0.14\%$ for $s_{W}^{2}$.
For $e+{}^{12}$C scattering, the achievable uncertainty
is
\begin{equation}
\frac{\Delta A^{{\rm PV}}}{A^{{\rm PV}}} \ \approx \ 0.3\% \, ,\ \
(e+^{12}{\rm C} \text{ scattering})
.\label{eq:m-18-1}
\end{equation}
Since $A^{{\rm PV}}(^{12}{\rm C})$ is proportional to $s_{W}^{2}$
rather than $1-4s_{W}^{2}$, the relative uncertainty for $s_{W}^{2}$
is also $0.3\%$.

With the experimental setup described above and the new physics contributions
formulated in Eqs.~\eqref{eq:m-12} and (\ref{eq:m-5}) to (\ref{eq:APV_epsC12}), it is straightforward to study the sensitivity
of P2 to $Z'$ models. By requiring that the new physics contributions do
not exceed
\begin{eqnarray}
\label{eq:factors}
\frac{\Delta A^{{\rm PV}}}{A^{{\rm PV}}} \ =
\ \begin{cases}
\sqrt{2.71}\times1.4\% \ = \ 2.30\% & (\text{for $e+p$ scattering}) \,, \\
\sqrt{2.71}\times0.3\% \ = \ 0.49\% & (\text{for $e+^{12}$C scattering}) \,,
\end{cases}
\end{eqnarray}
where the factor $\sqrt{2.71}$ converts 1$\sigma$ to 90\% confidence
level (C.L.), we obtain the P2 sensitivity curves in Figs.~\ref{fig:gx} to \ref{fig:U1B-L} for the variety of $U(1)'$ models under study.

As for the value of the weak mixing angle $\theta_W$ used in our numerical calculations, it is precisely measured at the electroweak scale from  $Z$ pole observables, the $W$ boson mass and a variety of NC processes. In the global fitting of high-energy data, it is determined to be $s^2_W =  0.23122 \pm 0.00003$~\cite{Zyla:2020zbs},
which can be run down via the SM renormalization group equations to lower energies~\cite{Erler:2004in, Erler:2017knj} and compared to the direct measurements of the weak mixing angle at that low-energy scales~\cite{Kumar:2013yoa}. In the numerical calculations, we have fixed the value of $s_W^2$ to its expected central low-$q^2$ value of $0.230 \pm 0.00003$.
As shown in Eq.\ (\ref{eq:m-2}), in the SM the parity-violating asymmetry $A^{\rm PV}_{\rm SM}$ in electron-proton scattering is proportional to the factor of $(1-4 s_W^2)$, therefore the uncertainty of $A^{\rm PV}_{\rm SM}$ due to the error bars $\Delta s_W^2$ of $s_W^2$ goes like
\begin{eqnarray}
\Delta A_{\rm SM}^{\rm PV} (e + p) \ \propto \ \frac{-4 \Delta s_W^2}{1-4 s_W^2} \,.
\end{eqnarray}
For scattering with $^{12}$C, the SM $A^{\rm PV}$ depends linearly on the weak mixing angle, i.e.
\begin{eqnarray}
\Delta A_{\rm SM}^{\rm PV} (e + ^{12}{\rm\! C}) \ \propto \ \frac{\Delta s_W^2}{s_W^2} \, .
\end{eqnarray}
In the case of $Z$-$Z'$ mixing, the new physics contributions to  $A^{\rm PV}$ are also subject to the weak mixing angle uncertainties, as implied in Eqs.~(\ref{eq:m-14}) to (\ref{eq:m-15}). They enter the result in the form of $\epsilon\, \Delta s_W^2$ or $\sin\theta \, \Delta s_W^2$. In light of the small experimental uncertainties of $s_W^2$, we neglect this higher-order effect  in this paper. 

Since the measurable $A^{\rm PV}$ is a single number, there is an unavoidable degeneracy between new physics and a  variation of $s_W^2$. This could, however, be resolved by performing the P2 measurements not only of the total asymmetry but of the angular distribution of the cross-sections over the whole range of scattering angle $\theta_f$ (see Eq.~(\ref{eq:m-17})).
In this way one could perform measurements at different $Q^2$, see Eq.\ (\ref{eq:Q2}), which could for appropriate $Z'$ masses break the degeneracy of the Weinberg angle with  new physics contributions to the asymmetry. Of course, measuring at two $Q^2$ with the same amount of available beam-time will cost precision.
In case interesting departures from the expectation are found during the run-time of the experiment, a trade-off between precision and new physics sensitivity could be a possibility. We note that a measurement at large scattering angle is foreseen by P2, in order to measure corrections from the proton structure to the measurement \cite{Becker:2018ggl}.






\subsection{Current experimental limits}
\label{sec:limits}

Before discussing our sensitivity results for P2, we briefly review the existing bounds on $Z'$ in each of the $U(1)'$ models considered in Sec.~\ref{sec:np}, following the compilations in,~e.g.~Refs.~\cite{Essig:2013lka,Ilten:2018crw,Bauer:2018onh,Berryman:2019dme, Fabbrichesi:2020wbt,Lanfranchi:2020crw}.
 Since the P2 experiment is based on electron-nucleon scattering,
bounds on electron and quark couplings should be taken into account.
In addition, for specific models, bounds on neutrino or muon couplings can also be used to constrain $g'$, $\epsilon$, and $\theta$. In addition, gauge invariance links electron scattering to neutrino scattering.

\newcommand{\subtitle}[1]{\vspace{3mm} \noindent $\bullet$ \textbf{#1} \vspace{1mm}}

\subtitle{Atomic parity violation:}

Precise low-energy measurements of APV have been performed using atomic $^{133}$Cs~\cite{Wood:1997zq, Bennett:1999pd, Porsev:2009pr, Porsev:2010de}, $^{205}$Tl~\cite{Edwards:1995zz, Vetter:1995vf}, $^{208}$Pb~\cite{Meekhof:1993zz},  $^{209}$Bi~\cite{Macpherson:1991opp} and $^{100,102,104,106}$Yb~\cite{Antypas:2018mxf}, and the most precise ones are the 6$S_{1/2}$ $-$ 7$S_{1/2}$ nuclear transition
in $^{133}$Cs. To lowest order, the nuclear weak charge measured in APV is given by $Q_W = - {\cal N}_{\rm Cs} + {\cal Z}_{\rm Cs} ( 1- 4s_W^2)$, with ${\cal Z}_{\rm Cs} = 55$ and ${\cal N}_{\rm Cs} = 78$ being the proton and neutron numbers in $^{133}$Cs respectively. Combining calculation of the nuclear spin-independent parity-violating electric dipole transition amplitude and the most accurate experimental measurements~\cite{Marciano:1982mm, Marciano:1983ss, Marciano:1990dp, Sahoo:2021thl}, we have the following SM predictions and experimental values of $Q_W ( ^{133}{\rm Cs} )$:
\begin{eqnarray}
Q_W^{\rm SM} ( ^{133}{\rm Cs} ) & \ = \ & -73.23(1) \,, \qquad
Q_W^{\rm exp} ( ^{133}{\rm Cs} )  \ = \  -73.71(35) \,,
\end{eqnarray}
which corresponds to an accuracy of $0.47\%$ ($0.77\%$) at the $1\sigma$ (90\%) C.L., and can be used to set limits on any new physics contributions to APV~\cite{Davoudiasl:2012ag, Safronova:2017xyt, Davoudiasl:2014kua, Abdullah:2018ykz}. In presence of the $Z'$ boson, the 6$S_{1/2}$ $-$ 7$S_{1/2}$ nuclear transition can be viewed as an electron-nucleus scattering in the Cesium atom, and the calculation methods for proton and $^{12}$C can be applied directly to the case of $^{133}$Cs, for instance Eqs.~(\ref{eq:m-12}), (\ref{eq:m-5}) and (\ref{eq:n-1}). For concreteness, we set the energy scale of $^{133}$Cs experiments to be $Q = 30$ MeV~\cite{Abdullah:2018ykz}, and the resultant limits are shown in Figs.~\ref{fig:gx} to \ref{fig:U1B-L} as the pink curves.

\subtitle{Beam dumps:}

Beam dump experiments such as E137~\cite{Bjorken:1988as}, E141~\cite{Riordan:1987aw},
E774~\cite{Bross:1989mp} and Orsay~\cite{Davier:1989wz} have searched for
 particles that could be produced in electron scattering off nuclei
in fixed targets and their subsequent decay to visible final states. In these experiments, the production process is electron bremsstrahlung  and the detection relies on $Z'\rightarrow e^{+}e^{-}$ decay. Therefore the searches are sensitive to the electron coupling of a low-mass $Z'$ in the range of 1 to 100 MeV. We use the package {\tt DARKCAST}~\cite{Ilten:2018crw} to recast and combine the bounds from these experiments.

There was recently a search of $e N \to eN X$ at NA64, with the $X$ boson decaying  invisibly~\cite{Andreev:2021xpu}. The null result at this experiment (which is not yet incorporated in {\tt DARKCAST}) sets an upper bound on the effective coupling of $X$ to electron, which can go up to the order of $10^{-5}$ for a MeV-scale $X$ boson mass and up to $10^{-2}$ if $X$ is at the GeV-scale, depending on the Lorentz structure of the couplings. When the NA64 limits in Ref.~\cite{Andreev:2021xpu} are cast onto the chiral and non-chiral $U(1)'$ models in this paper, the corresponding limits are expected to be the same order as that of the electron $g-2$ limits or up to one order of magnitude better (cf.\ Fig.~4 in Ref.~\cite{Andreev:2021xpu}), and are thus not shown in Figs.~\ref{fig:gx} to \ref{fig:U1B-L}.

\subtitle{Collider searches:}

New $Z'$ gauge bosons could be probed at colliders in several different ways.
For low-mass $Z'$ at $e^{+}e^{-}$ colliders, the mono-$\gamma$
channel $e^{+}e^{-}\rightarrow\gamma Z'$, followed by decays $Z'\rightarrow e^{+}e^{-}$ or $\mu^{+}\mu^{-}$ or into invisible states, is usually considered as
the most restrictive one. Searches in these channels have been performed
at BaBar~\cite{Lees:2014xha,Lees:2017lec}, NA48/2~\cite{Batley:2015lha},
KLOE~\cite{Babusci:2012cr} and LEP~\cite{Fox:2011fx}. At hadron
colliders, the Drell-Yan process $pp\rightarrow Z'\rightarrow\mu^{+}\mu^{-}$
or dijet, searched for by LHCb~\cite{Aaij:2019bvg}, ATLAS~\cite{Aad:2019fac, ATLAS:2019bov}, and CMS~\cite{CMS:2019tbu, Sirunyan:2019vgj}, produces the leading constraints
on heavy $Z'$ above the $Z$ pole. Besides, the LEP data can also
be used to constrain four-lepton effective operators generated by
heavy $Z'$~\cite{Abdallah:2005ph}. We adopt the ATLAS/CMS dijet
limit from Fig.~3 in Ref.~\cite{Bagnaschi:2019djj}.   Other collider
bounds are obtained using the {\tt DARKCAST} package~\cite{Ilten:2018crw}.
The LEP bounds in Figs.~\ref{fig:gx} and \ref{fig:U1B-L} (absent
in Figs.~\ref{fig:theta} to \ref{fig:U1B} due to insignificance) are
derived by fitting four-lepton effective operators to precision measurements
of $e^{+}e^{-}\rightarrow\ell^{+}\ell^{-}$ at LEP~\cite{Abdallah:2005ph}.\footnote{The LEP limits on flavored $Z'$ bosons can be found e.g. in Ref.~\cite{Chun:2018ibr}.}

\subtitle{Neutrino scattering:}

Elastic neutrino scattering data collected by CHARM II~\cite{Vilain:1993kd,Vilain:1994qy},
LSND~\cite{Auerbach:2001wg}, Borexino~\cite{Bellini:2011rx}, TEXONO~\cite{Deniz:2009mu} and
COHERENT~\cite{Akimov:2017ade} are sensitive
to a low-mass $Z'$ as it could mediate a $t$-channel process in $\nu+e^{-}$
or $\nu+N$ scattering enhanced by the low mass and low momentum transfer.
According to studies in Refs.~\cite{Bilmis:2015lja,Farzan:2018gtr,Lindner:2018kjo,Abdullah:2018ykz},
we select two of the most restrictive sets of data,  CHARM II and
TEXONO, and perform an independent data fitting\footnote{Code is available from \url{https://github.com/xunjiexu/Dark_Z}.}
for the various models considered in this work. The neutrino scattering bounds
apply to most of the models except for $U(1)_{R}$ and $U(1)_{B}$, in which left-handed neutrinos $\nu_{L}$ are not charged under the new $U(1)$. However, in the presence of kinetic/mass
mixing, these models could still be constrained by neutrino scattering.

\subtitle{Lepton anomalous magnetic moment:}

The muon and electron anomalous magnetic moments are sensitive to generic neutral bosons coupled to them.
The $Z'$ contribution to $a_{\ell}\equiv(g-2)/2$ can be evaluated
by~\cite{Leveille:1977rc}:
\begin{equation}
\Delta a_{\ell} \ = \ \frac{1}{8\pi^{2}}(g'_{\ell}\epsilon_{Z'})^{2}\int_{0}^{1}\frac{2x^{2}(1-x)}{(1-x)\left(1-x\epsilon_{Z'}^{2}\right)+x\epsilon_{Z'}^{2}}dx \, ,\label{eq:m-19}
\end{equation}
where $\epsilon_{Z'}\equiv m_{\ell}/m_{Z'}$ and $\ell=\mu$ or $e$.
The muon $g-2$ has a long-standing $3.7\sigma$ discrepancy between
the experimental and the calculated SM values~\cite{Aoyama:2020ynm}, i.e.
\begin{eqnarray}
\Delta a_\mu \ \equiv \ a_{\mu}^{{\rm exp}}-a_{\mu}^{{\rm SM}} \ = \
(27.9\pm 7.6)\times10^{-10} \,.
\label{eq:delta-mu}
\end{eqnarray}
The electron $g-2$ had earlier seen a $2.4\sigma$ discrepancy in the opposite direction~\cite{Davoudiasl:2018fbb}, based on the comparison between the SM prediction~\cite{Aoyama:2017uqe} and a measurement of the fine-structure constant using Cesium~\cite{Parker:2018vye}:
\begin{eqnarray}
\Delta a_e^{\rm Cs} \ \equiv \ a_{e}^{{\rm exp \ (Cs)}}-a_{e}^{{\rm SM}} \ = \ (-8.7\pm 3.6)\times10^{-13} \,.
\label{eq:deltae-Cs}
\end{eqnarray}
However, a recent measurement of the fine-structure constant using Rubidium~\cite{Morel:2020dww} has pushed the discrepancy to a mere $1.6\sigma$ in the same direction as $(g-2)_\mu$:
\begin{eqnarray}
\Delta a_e^{\rm Rb} \ \equiv \ a_{e}^{{\rm exp \ (Rb)}}-a_{e}^{{\rm SM}} \ = \ (4.8\pm 3.0)\times10^{-13} \,.
\end{eqnarray}
Although this weakens the possibility of any new physics contribution to $\Delta a_e$, the Cs and Rb measurements of $\alpha$ now disagree by more than $5\sigma$, which makes this whole issue  quite murky.
Here we take conservative bounds from $(g-2)_{\mu}$ and $(g-2)_{e}$
by requiring that $\Delta a_{\ell}$ does not exceed the experimental
best-fit values given by Eqs.~\eqref{eq:delta-mu} and \eqref{eq:deltae-Cs} by 5$\sigma$.  These bounds are presented in Figs.~\ref{fig:gx} to \ref{fig:U1B-L}
as green curves. Note that a $Z'$ that couples to electrons and muons with the same strength, as the case for the models studied here, can not explain the muon $g-2$ discrepancy~\cite{Lindner:2016bgg}.

\subtitle{ Electroweak precision tests:} \\
Electroweak precision tests can also constrain $Z'$ parameters. In Fig.\ \ref{fig:eps} we present a constraint on $\epsilon$ taken  from Ref.~\cite{Curtin:2014cca}. In principle, similar constraints are present for all the $Z'$ models, but need to be evaluated model-by-model, which is beyond the scope of our paper.

\subtitle{Loop-level meson and $Z$ decay limits} \\
The $Z'$ boson in the $U(1)'$ model could induce 1-loop level flavor-changing neutral current (FCNC) meson decays such as $B\rightarrow K +Z'$ and $K\rightarrow \pi +Z'$. The corresponding partial decay widths are not only directly relevant to the mass $m_{Z'}$, gauge coupling $g'$ (and also $\sin\theta$), but also depend subtly on the charges of the SM fermions under the $U(1)'$. For the $U(1)_{B-L}$ case, all the SM mass and interaction terms are  invariant under the $U(1)_{B-L}$ gauge symmetry, and the widths $\Gamma (K \to\pi Z')$ and $\Gamma (B \to K Z')$ can be found e.g. in Refs.~\cite{Pospelov:2008zw, Davoudiasl:2014kua}, which go to zero in the $m_{Z'}\to 0$ limit, as expected from angular momentum conservation. As for the other $U(1)'$ and $Z-Z'$ mixing models, the SM fermion couplings are explicitly chiral and so the fermion mass terms are not invariant under the $U(1)'$ gauge transformations. This leads to loop-level FCNC decays enhanced by the ratio $M^2/m_{Z'}^2$ for a very light $Z'$, with $M$ being the corresponding energy scale (typically the electroweak scale) for these decays~\cite{Dror:2017ehi, Dror:2017nsg, Dror:2018wfl}.  Similarly, one can also have the 1-loop level decay $\Upsilon \to \gamma Z'$ which is enhanced by $m^2_b/m^2_{Z'}$ when $Z'$ is light~\cite{Carone:1994aa, Fayet:2006sp, Fayet:2007ua}. Thus the searches of $\Upsilon \to \gamma + {\rm inv.}$~\cite{delAmoSanchez:2010ac} and $\Upsilon \to \gamma + X$ with $X \to \mu^+ \mu^-$~\cite{Love:2008aa} can be used to set limits on these $U(1)'$ models. If kinematically allowed, there is also the loop-level exotic $Z$ decay $Z \to \gamma Z'$. The searches of single photon from $Z$ decay~\cite{Acciarri:1997im, Abreu:1996pa} and $Z \to \gamma \ell^+ \ell^-$~\cite{Adriani:1992zm, Acton:1991dq, Adeva:1991dw} at the LEP can be used to constrain the $Z'$ bosons. However,  these FCNC decay constraints become important only for sub-MeV scale $Z'$ bosons, and are not relevant for our P2 sensitivity analysis. Moreover, the $M^2/m_Z'^2$-enhanced bounds cannot be arbitrarily strong in the $m_{Z'}\to 0$ limit and one has to take the full momentum dependence into account in loop calculations.


\subtitle{ Astrophysical and cosmological limits:} \\
A sufficiently light $Z'$ can also be constrained by astrophysical limits such
as from supernova 1987A~\cite{Rrapaj:2015wgs, Chang:2016ntp} and cosmological limits such as from the big bang nucleosynthesis (BBN)~\cite{Fradette:2014sza, Escudero:2019gzq}. However, these limits mostly lie outside the window of parameter space considered in Figs.~\ref{fig:gx} to \ref{fig:U1B-L}. The BBN limits are significant only for $Z'$ masses below 10 MeV, and the supernova limits apply additionally only for small couplings, see e.g.\ Ref.~\cite{Knapen:2017xzo}. Moreover, the BBN bound depends on the branching ratio of $Z'\rightarrow\nu\overline{\nu}$ and $Z'\rightarrow e^{-}e^{+}$.  Thus these astrophysical and cosmological limits are not relevant to the P2 prospects, and thus are not shown in Figs.~\ref{fig:gx} to \ref{fig:U1B-L}.

\subsection{P2 sensitivities}
\label{sec:sensitivities}
In this subsection, we derive the P2 sensitivities for the models considered in Sec.~\ref{sec:np} and compare them with the current experimental constraints discussed above.
\subsubsection{Chiral models}

\begin{figure}[t]
\centering
\includegraphics[width=0.8\textwidth]{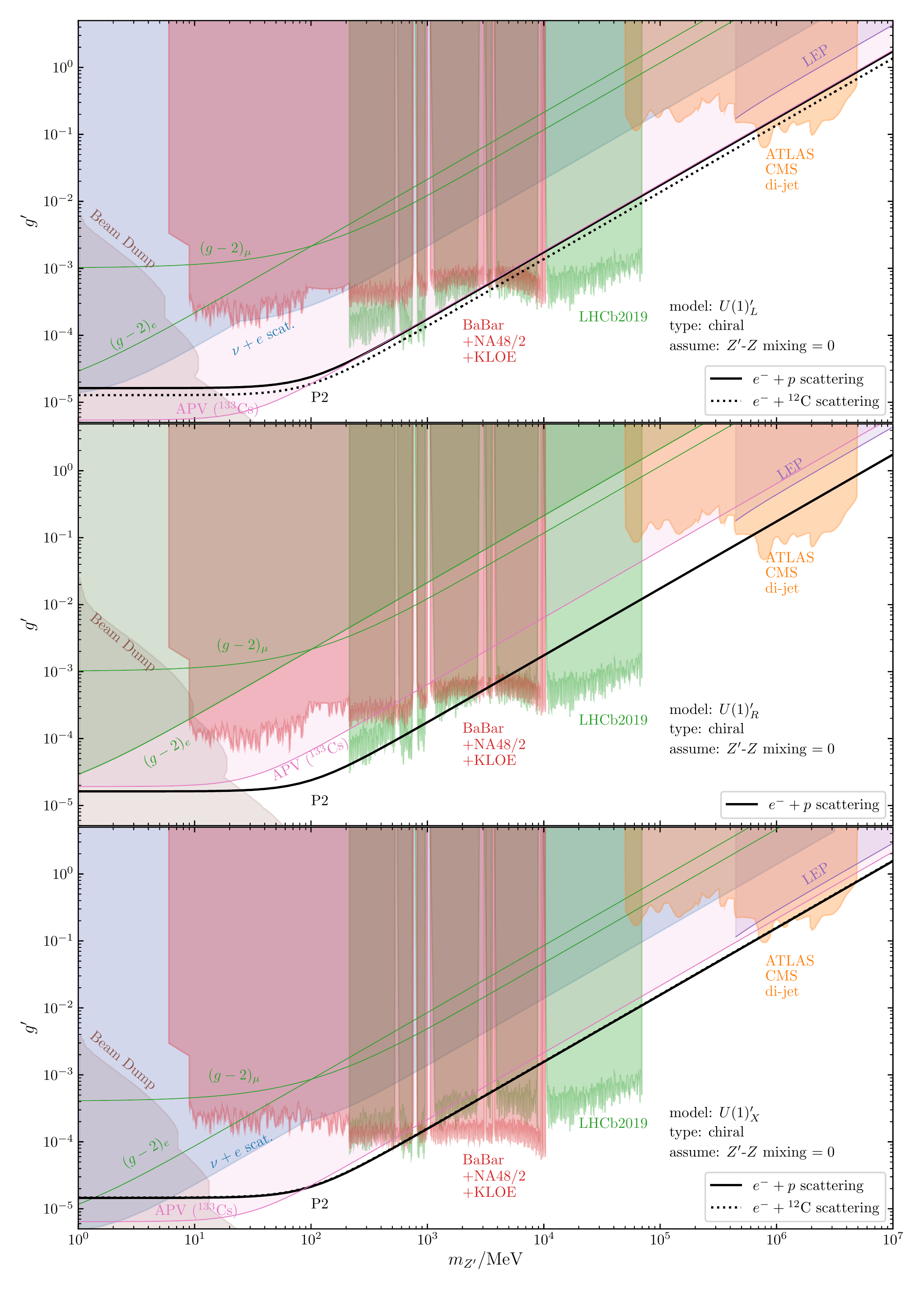}
\caption{Sensitivities of the P2 experiment in $e+ p$ (solid black) and $e+ ^{12}$C (dashed black) scatterings to the gauge coupling $g'$ in three
chiral $U(1)'$ models given in Tab.~\ref{tab:chiral_models}. For simplicity we have assumed here that there is no mass or kinetic $Z$-$Z'$ mixing. The shaded regions are excluded by  APV measurements~\cite{Wood:1997zq, Bennett:1999pd, Porsev:2009pr, Porsev:2010de}, beam-dump experiments~\cite{Bjorken:1988as, Riordan:1987aw, Bross:1989mp, Davier:1989wz}, electron and muon $g-2$~\cite{Zyla:2020zbs, Davoudiasl:2018fbb}, BaBar~\cite{Lees:2014xha,Lees:2017lec}, NA48/2~\cite{Batley:2015lha},
KLOE~\cite{Babusci:2012cr}, LHCb~\cite{Aaij:2019bvg}, LEP $e^+ e^- \to \ell^+ \ell^-$ data~\cite{Abdallah:2005ph} and the ATLAS and CMS dijet data~\cite{Bagnaschi:2019djj}. Note that, due to the cancellation of quantum numbers in Eq.~(\ref{eq:m-11}), the P2 prospect in the $e+ ^{12}$C channel is absent for the $U(1)'_R$ model (middle panel). See Sections~\ref{sec:limits} and \ref{sec:sensitivities} for more details.
\label{fig:gx}}
\end{figure}

The P2 sensitivities for the three chiral $U(1)'$ models given in Table~\ref{tab:chiral_models} are shown in Fig.~\ref{fig:gx}, where the $e+p$ sensitivities are shown as solid black lines, while the $e + ^{12}$C sensitivities are in dashed black lines. All existing relevant limits discussed in Section~\ref{sec:limits} are presented in Fig.~\ref{fig:gx} as the shaded regions. Here we have assumed that there is no mass or kinetic $Z-Z'$ mixing. It is clear from Fig.~\ref{fig:gx} that the limits for the three chiral $U(1)'$ models are rather similar: at low masses $m_{Z'} \lesssim 100$ GeV, the strongest limits are mainly from APV measurements (see below), beam-dump experiments, and from BaBar, NA48/2 and KLOE  and LHCb  data. For the models $U(1)_L^\prime$ and $U(1)_X^\prime$ the limits from $\nu+e$ scattering are also important,  while for $U(1)'_R$ the $Z'$ boson does not couple directly to neutrinos (see Table~\ref{tab:chiral_models}), and as a result these are no neutrino scattering limits for the $U(1)'_R$ model. Suppressed by the charged lepton masses (see Eq.~(\ref{eq:m-19})), the electron and muon $g-2$ limits are relatively weaker for  these chiral models (and also for all other models in this paper).
When $Z'$ is heavier than the electroweak scale, the coupling $g'$ is constrained by the direct searches of $Z'$ in the LHC dijet data and the limits from the LEP $e^+ e^- \to \ell^+ \ell^-$ data.

Comparing the solid and dashed black lines in the upper panel of Fig.~\ref{fig:gx}, we see that the $e + ^{12}$C sensitivity for the $U(1)'_L$ model is more stringent than that from $e + p$ scattering by a factor of 1.4. This is mainly due to the enhancement of the effective coupling of $Z'$ to the $^{12}$C nucleus with respect to the proton. In particular, following Eqs.~(\ref{eq:m-10}) and (\ref{eq:m-11}), we have
\begin{eqnarray}
Q'_p  \ = \ \frac12 \,, \quad
Q'_{^{12}{\rm C}}  \ = \ 12 \, \quad \text{for $U(1)'_L$} \,.
\end{eqnarray}
Taking into account the two different factors for $\Delta A^{\rm PV} / A^{\rm PV}_{\rm SM}$ in Eq.~(\ref{eq:factors}),  and the SM predictions of $A^{\rm PV}_{\rm SM}$ for proton and $^{12}$C in Eqs.~(\ref{eq:m-2}) and (\ref{eq:m-2-1}), one gets the enhancement factor of 1.4. Similarly, as a result of
\begin{eqnarray}
Q'_p  \ = \ \frac54 \,, \quad
Q'_{^{12}{\rm C}}  \ = \ 18 \,  \quad \text{for $U(1)'_X$} \,,
\end{eqnarray}
the  $e + p$ and $e + ^{12}$C sensitivities for the $U(1)'_X$ model are almost the same, as can be seen in the bottom panel of Fig.~\ref{fig:gx}. For the $U(1)'_R$ model, due to the accidental cancellation of
\begin{eqnarray}
Q'_{u_L}+Q'_{u_R} + Q'_{d_L}+Q'_{d_R} \ = \ 0 \,,
\label{eq:acc}
\end{eqnarray}
we find that $Q'_{^{12}{\rm C}}  = 0$, and there is no $e + ^{12}$C sensitivity for the $U(1)'_R$ model.

\begin{table}
  \centering
  \caption{P2 sensitivities for the chiral and non-chiral models shown in Figs.~\ref{fig:gx} to \ref{fig:U1B-L} in the limits of $m_{Z'} \to 0$ and $m_{Z'} \to \infty$. ``$U(1)$ ($\sin\theta$, $\epsilon$)'' refers to the generic $U(1)'$ model with only mass mixing $\sin\theta$ or kinetic mixing $\epsilon$. The numbers in the parentheses for the $U(1)_B$ and $U(1)_{B-L}$ models are the values of $g'/\sin\theta$. Note that for the $U(1)'_R$ model there is no sensitivity in the  $e + ^{12}$C scattering mode due to an accidental cancellation (cf.~Eq.~\eqref{eq:acc}).
  \vspace{-15pt}
  \label{tab:limits}}
  \scalebox{0.92}{
  \begin{tabular}{c|c|c|c|c}
\hline\hline
\multirow{2}{*}{models} & \multicolumn{2}{c|}{$m_{Z'}\to0$} & \multicolumn{2}{c}{$m_{Z'}\to \infty$} \\ \cline{2-5}
 & $e + p$  &   $e + ^{12}$C & $e + p$  &   $e + ^{12}$C \\ \hline
$U(1)_{L}'$ & $g' < 1.7\times10^{-5}$ & $g' < 1.2\times10^{-5}$ & $\frac{m_{Z'}}{g'} > 5.9$ TeV & $\frac{m_{Z'}}{g'} > 8.3$ TeV \\ \hline
$U(1)_{R}'$ & $g' < 1.7\times10^{-5}$ & $-$ & $\frac{m_{Z'}}{g'} > 5.6$ TeV & $-$ \\ \hline
$U(1)_{X}'$ & $g' < 1.5\times10^{-5}$ & $g' < 1.5\times10^{-5}$ & $\frac{m_{Z'}}{g'} > 7.7$ TeV & $\frac{m_{Z'}}{g'} > 7.7$ TeV \\ \hline
$U(1)'$ ($\sin\theta$) & $\sin\theta < 1.7\times10^{-4}$ & $\sin\theta < 7.3\times10^{-5}$ & $\sin\theta < 0.13$ & $\sin\theta < 0.07$ \\ \hline
$U(1)'$ ($\epsilon$) & $\epsilon < 0.05$ & $\epsilon < 0.07$ & $\frac{m_{Z'}}{\epsilon} > 350$ GeV & $\frac{m_{Z'}}{\epsilon} > 600$ GeV \\ \hline
$U(1)_B$ (0.01) & $\sin\theta < 9\times10^{-5}$ & $\sin\theta < 9\times10^{-5}$ & $\sin\theta < 0.13$ & $\sin\theta < 0.07$ \\ \hline
$U(1)_B$ (1) & $\sin\theta < 1.1\times10^{-5}$ & $\sin\theta < 1.2\times10^{-5}$ & $\sin\theta < 0.13$ & $\sin\theta < 0.07$ \\ \hline
$U(1)_B$ (10) & $\sin\theta < 3.7\times10^{-6}$ & $\sin\theta < 4.1\times10^{-6}$ & $\sin\theta < 0.13$ & $\sin\theta < 0.07$ \\ \hline
$U(1)_{B-L}$ (0.01) & $\sin\theta < 2.9\times10^{-4}$ & $\sin\theta < 6.9\times10^{-5}$ & $\sin\theta < 0.13$ & $\sin\theta < 0.07$ \\ \hline
$U(1)_{B-L}$ (1) & $\sin\theta < 2\times10^{-5}$ & $\sin\theta < 2\times10^{-5}$ & $\sin\theta < 0.13$ & $\sin\theta < 0.07$ \\ \hline
$U(1)_{B-L}$ (10) & $\sin\theta < 6\times10^{-6}$ & $\sin\theta < 6.5\times10^{-6}$ & $\sin\theta < 0.13$ & $\sin\theta < 0.07$ \\
\hline\hline
\end{tabular}}
\end{table}

\begin{table}
    \centering
    \caption{
    Ranges of $m_{Z'}$ and $g'$ where P2 will be able to improve current limits for chiral $U(1)'$ models. This table is a summary of Fig.~\ref{fig:gx}. The columns of $U(1)'_L$ in the $e+p$ mode are void because P2 does not have sensitivity beyond existing APV limits.
    \label{tab:limits2a}}
\begin{tabular}{c|c|c|c}
\hline\hline
models & mode & $m_{Z'}$ ranges & $g'$ ranges  \\ \hline
\multirow{2}{*}{$U(1)_{L}'$} & $e + p$ & $-$ & $-$
  \\ \cline{2-4}
& $e + ^{12}$C
  & \begin{tabular}{@{}c@{}} $\text{[100 MeV, 200 MeV]}$ \\ $\text{[70 GeV, 600 GeV]}$ \\ $\text{[5 TeV, 90 TeV]}$ \end{tabular} &
  \begin{tabular}{@{}c@{}} $[2.0 \times 10^{-5},\, 3.3 \times 10^{-5}]$ \\ $[0.01,\, 0.09]$ \\ $[0.70,\, 4\pi]$ \end{tabular} \\ \hline
$U(1)_{R}'$ & $e + p$
  & \begin{tabular}{@{}c@{}} $\text{[20 MeV, 200 MeV]}$ \\ $\text{[70 GeV, 600 GeV]}$ \\ $\text{[5 TeV, 70 TeV]}$ \end{tabular} &
  \begin{tabular}{@{}c@{}} $[1.7 \times 10^{-5},\, 4.0 \times 10^{-5}]$ \\ $[0.012,\, 0.1]$ \\ $[0.9,\, 4\pi]$ \end{tabular} \\ \hline
\multirow{2}{*}{$U(1)_{X}'$} & $e + p$
  & \begin{tabular}{@{}c@{}} $\text{[93 MeV, 430 MeV]}$ \\ $\text{[70 GeV, 700 GeV]}$ \\ $\text{[5 TeV, 79 TeV]}$ \end{tabular} &
  \begin{tabular}{@{}c@{}} $[2.0 \times 10^{-5},\, 7.5 \times 10^{-5}]$ \\ $[0.01,\, 0.11]$ \\ $[0.8,\, 4\pi]$ \end{tabular} \\ \cline{2-4}
& $e + ^{12}$C
  & \begin{tabular}{@{}c@{}} $\text{[93 MeV, 430 MeV]}$ \\ $\text{[70 GeV, 700 GeV]}$ \\ $\text{[5 TeV, 79 TeV]}$ \end{tabular} &
  \begin{tabular}{@{}c@{}} $[2.0 \times 10^{-5},\, 7.5 \times 10^{-5}]$ \\ $[0.01,\, 0.11]$ \\ $[0.8,\, 4\pi]$ \end{tabular} \\
\hline\hline
\end{tabular}
\end{table}

%

For masses $m_{Z'} \lesssim 100$ MeV, the P2 sensitivities are almost independent of the $Z'$ mass, which correspond to the plateau in the three panels of Fig.~\ref{fig:gx}. This is also true for the other models considered here (see  Figs.~\ref{fig:theta} to \ref{fig:U1B-L}). In the high $m_{Z'}$ limit, i.e.\ $m_{Z'} \to \infty$, we can safely neglect the momentum transfer $Q^2$, and the P2 experiment can probe an effective ultraviolet (UV) cutoff scale of $\Lambda = m_{Z'} /g'$ at the few-TeV scale. All the P2 sensitivities in the limits of $m_{Z'} \to 0$ and $m_{Z'} \to \infty$ for the three chiral models are collected in Table~\ref{tab:limits}.

The low-energy nuclear transition in $^{133}$Cs is essentially equivalent to electron-nucleus scattering, thus for all our cases the properties of the APV limits are very similar to those for the P2 prospects.  Comparing the $^{133}$Cs atom in the most precise APV measurement with $^{12}$C in P2, the differences lie mainly in the following factors: the nuclear weak charges, the effective couplings of $Z'$ to nuclei and the energy scale $Q$. Taking the $U(1)'_L$ model as an explicit example, the coupling of $Z'$ to $^{133}$Cs is
\begin{eqnarray}
Q'_{^{133}{\rm Cs}} \ = \ \frac{1}{2}
\left( 266  \, Q'_{u_L} + 266 \, Q'_{u_R} + 211 \, Q'_{d_L} + 211 \, Q'_{d_R} \right)
\ = \ \frac{263}{2} \,.
\end{eqnarray}
Following Eq.~(\ref{eq:m-12}), the resulting parity violation for the atom $X$ with respect to the SM value is proportional to
\begin{eqnarray}
\label{eqn:APV1}
\frac{ \Delta A^{\rm PV} (X)}{ A_{\rm SM}^{\rm PV} } \ \propto \
Q^{-2} \frac{Q'_{X}}{Q_W (X)}
\frac{Q^2}{Q^2 + m_{Z'}^2} \,.
\end{eqnarray}
In the limit of $m_{Z'} \ll Q$, we can neglect the last factor in the equation above, and compare the APV sensitivities in $^{133}$Cs and the P2 prospects using $^{12}$C:
\begin{eqnarray}
\label{eqn:APV:comparison}
\frac{ \Delta A^{\rm PV} ( ^{133}{\rm Cs} ) / A_{\rm SM}^{\rm PV} ( ^{133}{\rm Cs} ) } { \Delta A^{\rm PV} ( ^{12}{\rm C} ) / A_{\rm SM}^{\rm PV} ( ^{12}{\rm C} ) } (m_{Z'} \ll Q) & \ = \ &
\left( \frac{Q_{\rm P2}}{Q_{\rm APV}} \right)^2
\frac{ Q'_{^{133}{\rm Cs}} / Q_W (^{133}{\rm Cs}) }{ Q'_{^{12}{\rm C}} / Q_W (^{12}{\rm C}) } \nonumber \\
& \ \simeq \ &
\left( \frac{93\; {\rm MeV}}{30\; {\rm MeV}} \right)^2
\left( \frac{ -1.78 }{ -2.17 } \right) .
\end{eqnarray}
Taking into account the difference of the APV accuracy (0.77\%) and the P2 precision in $^{12}$C (0.49\%), Eq.~(\ref{eqn:APV:comparison}) implies that for the $U(1)'$ model at  low energies  $m_{Z'} \ll Q$, the APV limit on $g'$ is more stringent than the P2 prospect using $^{12}$C by a factor of
\begin{eqnarray}
\left( \frac{0.49\%}{0.77\%}  \right)^{1/2}
\left( \frac{93\; {\rm MeV}}{30\; {\rm MeV}}  \right)
\left( \frac{ -1.78 }{ -2.17 } \right)^{1/2} \ \simeq \ 2.24 \,.
\end{eqnarray}
In the limit of $m_{Z'} \gg Q$, the last factor in Eq.~(\ref{eqn:APV1}) can be simplified to  $Q^2/m_{Z'}$ and any momentum dependence cancels. In this case the comparison of APV  and P2  with the target $^{12}$C is determined only by the coupling factors $Q'_N$ and $Q_W$, i.e.
\begin{eqnarray}
\label{eqn:APV:comparison2}
\frac{ \Delta A^{\rm PV} ( ^{133}{\rm Cs} ) / A_{\rm SM}^{\rm PV} ( ^{133}{\rm Cs} ) } { \Delta A^{\rm PV} ( ^{12}{\rm C} ) / A_{\rm SM}^{\rm PV} ( ^{12}{\rm C} ) } (m_{Z'} \gg Q) & \ = \ &
\frac{ Q'_{^{133}{\rm Cs}} / Q_W (^{133}{\rm Cs}) }{ Q'_{^{12}{\rm C}} / Q_W (^{12}{\rm C}) }
 \ \simeq \ \left( \frac{ -1.78 }{ -2.17 } \right) ,
\end{eqnarray}
and the APV limit on $g'$ is weaker than the P2 prospect with $^{12}$C by a factor of
\begin{eqnarray}
\left( \frac{0.49\%}{0.77\%}  \right)^{1/2}
\left( \frac{ -1.78 }{ -2.17 } \right)^{1/2} \ \simeq \ 0.73 \,,
\end{eqnarray}
as shown in the upper panel of Fig.~\ref{fig:gx}. In a similar way, one can compare the APV limits with the P2 sensitivities for the proton target for the $U(1)'$ models. It turns out that for the $U(1)'_L$ model the APV constraint is almost the same as that for the P2 prospect with the proton target in the limit of large $m_{Z'}$. For the $U(1)'_R$ model, the APV limit is slightly weaker than the P2 sensitivity with protons for all values of $m_{Z'}$. For the $U(1)'_X$ model the  APV limits exclude the P2 prospects at the low mass range of $m_{Z'}$, and are slightly weaker than the P2 sensitivities for both proton and $^{12}$C when the $Z'$ mass is large.

The P2 sensitivities in the lower mass range have also been precluded partially by the beam-dump experiments, NA62, Na48/2 and KLOE experiments, and in the high mass range partially by the LHC dijet data. However, even if all these constraints are taken into consideration, there is still some parameter space in the $m_{Z'}$-$g'$ plane that can be probed at the P2 experiment, which is collected in Table~\ref{tab:limits2a}. Note that the P2 prospects for the $U(1)'_L$ model in the proton mode have been precluded by the APV limits. Depending on the $U(1)'$ model, $Z'$ boson masses can be probed down to roughly 20 MeV, and the coupling $g'$ can be probed down to the order of $10^{-5}$.
In the high $Z'$ mass end, if the gauge coupling is fixed at the perturbative limit of $4\pi$, the P2 experiment can reach $Z'$ masses up to 70 TeV and 79 TeV for the chiral $U(1)'_R$ and $U(1)'_X$ models respectively in $e+p$ scattering, and up to 90 TeV and 79 TeV for the $U(1)'_L$ and $U(1)'_X$ models in $e+^{12}$C mode. The P2 sensitivities of heavy $Z'$ bosons are largely complementary to the direct searches at the LHC and future higher energy colliders.





\subsubsection{Non-chiral models}

\begin{figure}[t]
\centering
\includegraphics[width=0.87\textwidth]{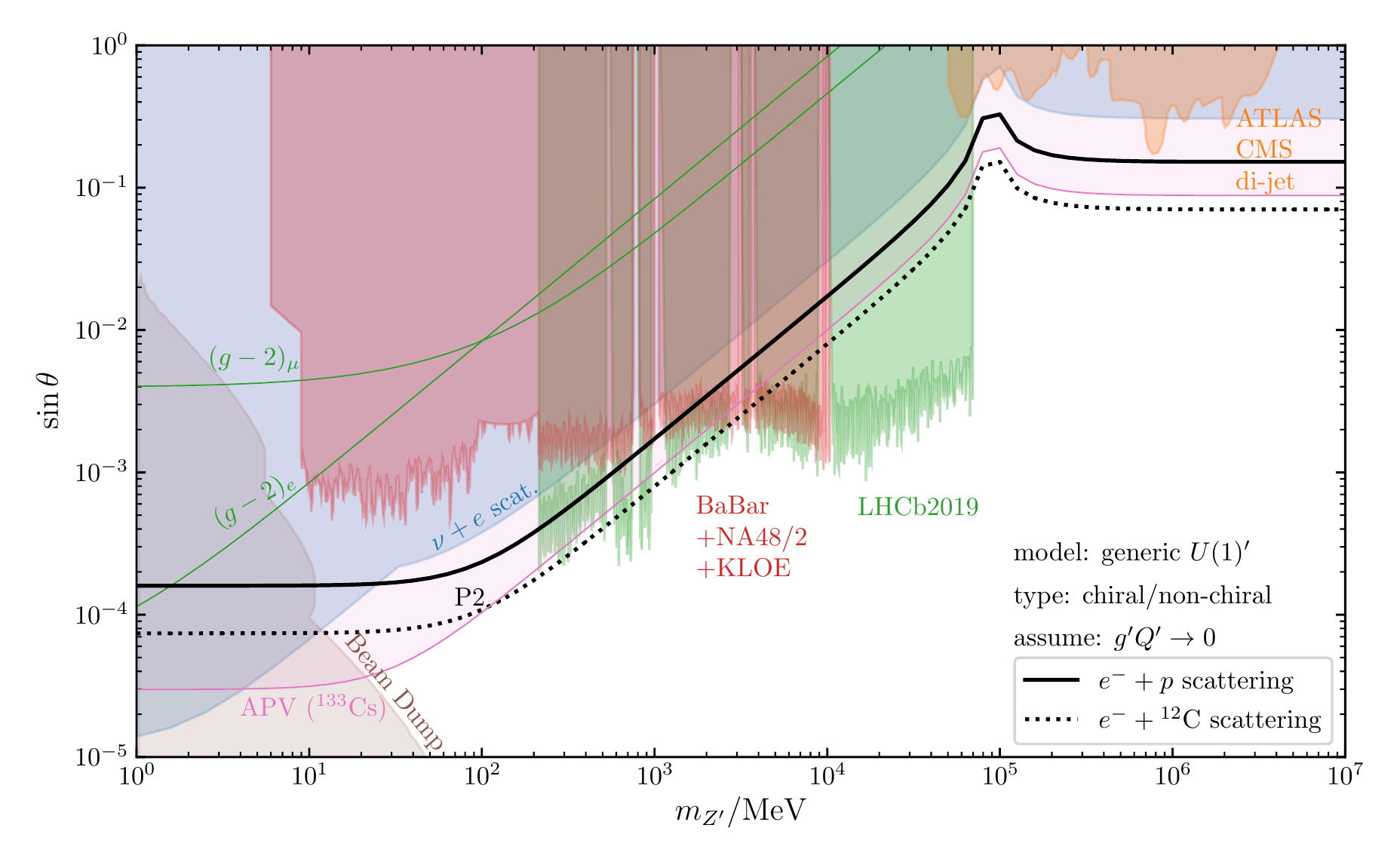}
\caption{Sensitivity of the P2 experiment to the generic $U(1)'$ model with only mass mixing $\sin\theta$ and in the limit of $g'Q'=0$. The notation is the same as in Fig.~\ref{fig:gx}.
\label{fig:theta}
}
\end{figure}

\begin{figure}[t]
\centering
\includegraphics[width=0.87\textwidth]{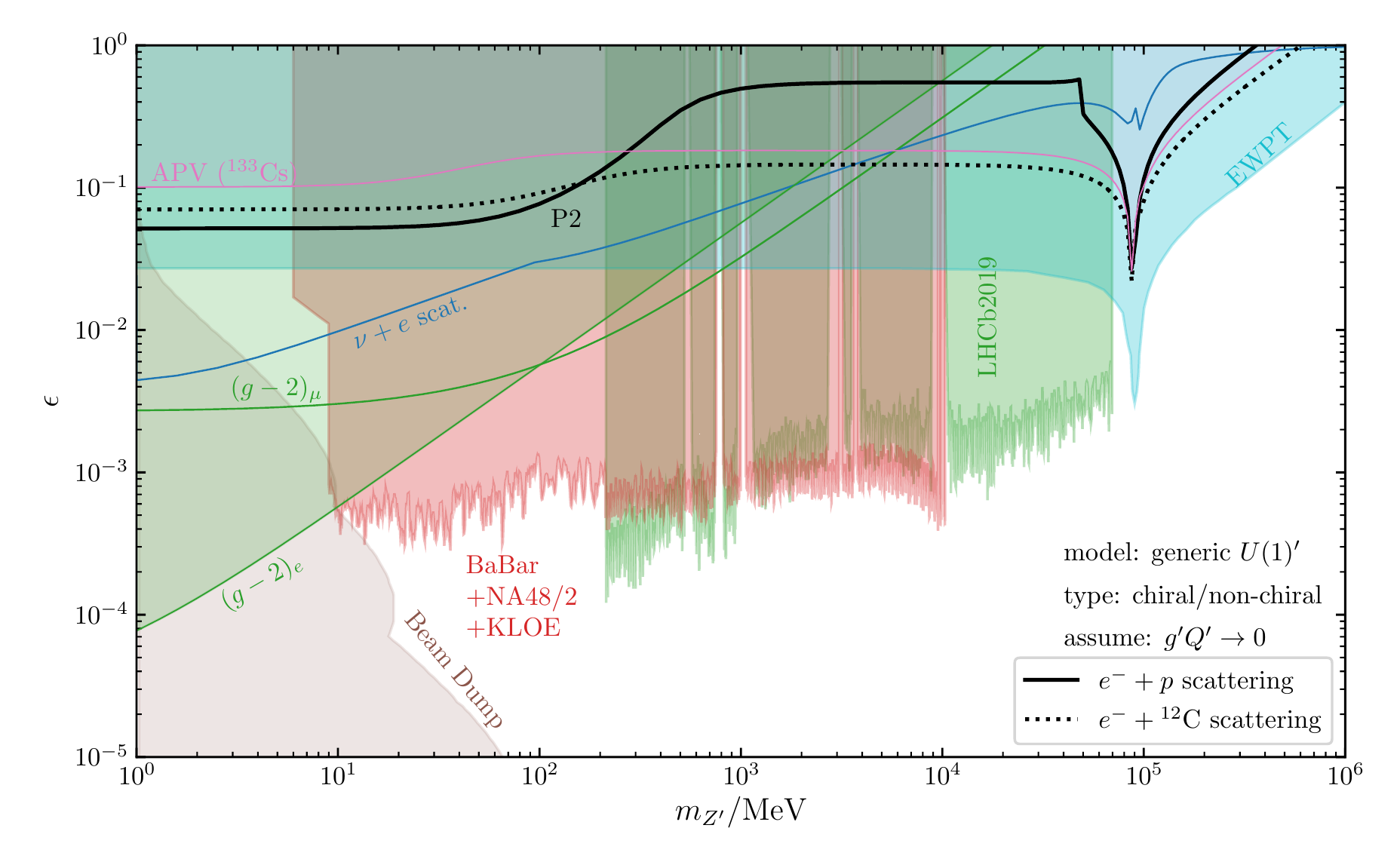}
\caption{Sensitivity of the P2 experiment to the generic $U(1)'$ model with only kinetic mixing $\epsilon$ and in the limit of $g'Q'=0$. The notation is the same as in Fig.~\ref{fig:gx}.
\label{fig:eps}}
\end{figure}

For the generic non-chiral $U(1)'$ model with $Z$-$Z'$ mass mixing, the P2 sensitivities and corresponding existing limits are shown in Fig.~\ref{fig:theta} for the simplest case of $g' Q' = 0$. This corresponds to the cases of negligibly small couplings ($g'/\sin\theta\ll1$) or no SM fermions carrying $U(1)'$ charges (known as the secluded $Z'$ model).
The limits from LEP $e^+ e^- \to \ell^+ \ell^-$ data are very weak and thus are not shown in Fig.~\ref{fig:theta}. The P2 sensitivity for this model is given in Eq.~(\ref{eq:m-5}). As in Fig.~\ref{fig:gx}, for  $m_{Z'} \lesssim$ 100 MeV, the P2 sensitivities are almost independent of $m_{Z'}$. In the mass range  $100 \; {\rm MeV} \lesssim m_{Z'} \lesssim m_Z$, the factor in the parentheses of Eq.~(\ref{eq:m-5}) can be simplified to be
\begin{eqnarray}
\label{eqn:factor}
\frac{m_{Z}^{2}}{m_{Z'}^{2}+Q^{2}}-1 \ \simeq \  \frac{m_{Z}^2}{m_{Z'}^2} \,.
\end{eqnarray}
Therefore, in this mass range P2 excludes an effective cutoff scale $\Lambda = m_{Z'} / \sin\theta$, which is 560 GeV and 1.1 TeV for the $e+ p$ and $e+ ^{12}$C scattering, respectively. The P2 sensitivity vanishes when the $Z'$ mass is close to $m_Z$. When $m_{Z'} > m_{Z}$, the factor in Eq.~(\ref{eqn:factor}) changes its sign, and approaches the value $-1$ in the limit of $m_{Z'} \gg m_{Z}$. As a result, the P2 sensitivities are constant in the heavy $Z'$ limits, which are respectively $\sin\theta < 0.13$ and $0.07$ for  $e+ p$ and $e+ ^{12}$C scattering.

\begin{table}
    \centering
    \caption{
      Similar to Tab.~\ref{tab:limits2a} except for non-chiral $U(1)'$ models, summarized from Figs.~\ref{fig:theta} to \ref{fig:U1B-L}.
      The columns of $U(1)'$ ($\sin\theta$) in the $e+p$ mode and $U(1)'$ ($\epsilon$) for all models are void because P2 does not have better sensitivities than the existing limits.
    \label{tab:limits2b}}
\begin{tabular}{c|c|c|c}
\hline\hline
models & mode & $m_{Z'}$ ranges & $\sin\theta$ or $\epsilon$ ranges  \\ \hline
\multirow{2}{*}{$U(1)'$ ($\sin\theta$)} & $e + p$ & $-$ & $-$
  \\ \cline{2-4}
& $e + ^{12}$C
  & \begin{tabular}{@{}c@{}} $\text{[110 MeV, 200 MeV]}$ \\ ($>70$ GeV) \end{tabular} &
  \begin{tabular}{@{}c@{}} $[1.1 \times 10^{-4},\, 1.9 \times 10^{-4}]$ \\  $[0.07,\, 0.15]$ \end{tabular} \\ \hline
\multirow{2}{*}{$U(1)'$ ($\epsilon$)} & $e + p$
  & $-$ & $-$ \\ \cline{2-4}
& $e + ^{12}$C
  & $-$ & $-$ \\ \hline
\multirow{2}{*}{$U(1)_B$ (0.01)} & $e + p$
  & \begin{tabular}{@{}c@{}} $\text{[120 MeV, 200 MeV]}$ \\ $\text{[70 GeV, 100 GeV]}$ \end{tabular} &
  \begin{tabular}{@{}c@{}} $[1.4 \times 10^{-4},\, 2.3 \times 10^{-4}]$ \\  $[0.08,\, 0.20]$ \end{tabular} \\ \cline{2-4}
& $e + ^{12}$C
  & \begin{tabular}{@{}c@{}} $\text{[120 MeV, 200 MeV]}$ \\ ($>70$ GeV) \end{tabular} &
  \begin{tabular}{@{}c@{}} $[1.4 \times 10^{-4},\, 2.3 \times 10^{-4}]$ \\  $[0.07,\, 0.13]$ \end{tabular} \\ \hline
\multirow{2}{*}{$U(1)_{B}$ (1)} & $e + p$
  & \begin{tabular}{@{}c@{}} $\text{[70 MeV, 10 GeV]}$ \\ $\text{[70 GeV, 500 GeV]}$ \end{tabular} &
  \begin{tabular}{@{}c@{}} $[1.2 \times 10^{-5},\, 1.2 \times 10^{-3}]$ \\  $[8 \times 10^{-3},\, 0.06]$ \end{tabular} \\ \cline{2-4}
& $e + ^{12}$C
  & \begin{tabular}{@{}c@{}} $\text{[83 MeV, 10 GeV]}$ \\ $\text{[70 GeV, 650 GeV]}$ \\ ($>2.3$ TeV) \end{tabular} &
  \begin{tabular}{@{}c@{}} $[1.4 \times 10^{-5},\, 1.3 \times 10^{-3}]$ \\  $[9 \times 10^{-3},\, 0.055]$ \\ $[0.067,\, 0.07]$ \end{tabular} \\ \hline
\multirow{2}{*}{$U(1)_{B}$ (10)} & $e + p$
  &  $\text{[70 MeV, 305 GeV]}$  & $[4.0 \times 10^{-6},\, 0.012]$   \\ \cline{2-4}
& $e + ^{12}$C
  & \begin{tabular}{@{}c@{}} $\text{[83 MeV, 305 GeV]}$ \\ ($>5$ TeV) \end{tabular} &
  \begin{tabular}{@{}c@{}} $[4.2 \times 10^{-6},\, 0.012]$ \\ $[0.067,\, 0.07]$ \end{tabular} \\ \hline
\multirow{2}{*}{$U(1)_{B-L}$ (0.01)} & $e + p$
  & [90 GeV, 110 GeV] & $[0.17,\, 0.19]$ \\ \cline{2-4}
& $e + ^{12}$C
  & \begin{tabular}{@{}c@{}} $\text{[110 MeV, 560 MeV]}$ \\ ($>70$ GeV) \end{tabular} &
  \begin{tabular}{@{}c@{}} $[1 \times 10^{-4},\, 5.0 \times 10^{-4}]$ \\ $[0.07,\, 0.2]$ \end{tabular} \\ \hline
\multirow{2}{*}{$U(1)_{B-L}$ (1)} & $e + p$
  & \begin{tabular}{@{}c@{}} $\text{[80 MeV, 600 MeV]}$ \\ $\text{[70 GeV, 620 GeV]}$ \end{tabular} &
  \begin{tabular}{@{}c@{}} $[2.4 \times 10^{-5},\, 1.2 \times 10^{-4}]$ \\ $[0.013,\, 0.1]$ \end{tabular} \\ \cline{2-4}
& $e + ^{12}$C
  & \begin{tabular}{@{}c@{}} $\text{[80 MeV, 600 MeV]}$ \\ $\text{[70 GeV, 260 GeV]}$ \\ ($>400$ GeV) \end{tabular} &
  \begin{tabular}{@{}c@{}} $[2.4 \times 10^{-5},\, 1.2 \times 10^{-4}]$ \\ $[0.013,\, 0.18]$ \end{tabular} \\ \hline
\multirow{2}{*}{$U(1)_{B-L}$ (10)} & $e + p$
  & \begin{tabular}{@{}c@{}} $\text{[70 MeV, 200 MeV]}$ \\ $\text{[70 GeV, 420 GeV]}$ \end{tabular} &
  \begin{tabular}{@{}c@{}} $[7.0 \times 10^{-6},\, 1.4 \times 10^{-5}]$ \\ $[4.3 \times 10^{-3},\, 0.028]$ \end{tabular} \\ \cline{2-4}
& $e + ^{12}$C
  & \begin{tabular}{@{}c@{}} $\text{[83 MeV, 600 MeV]}$ \\ $\text{[70 GeV, 420 GeV]}$ \end{tabular} &
  \begin{tabular}{@{}c@{}} $[9.0 \times 10^{-6},\, 1.5 \times 10^{-5}]$ \\ $[4.5 \times 10^{-3},\, 0.032]$ \end{tabular} \\
\hline\hline
\end{tabular}
\end{table}

Similar to the chiral models, the APV limits in the generic $U(1)'$ model have the same features as those for the P2 prospects. In the limit of $m_{Z'} \ll Q$, Eq.~(\ref{eq:m-5}) implies that the new physics contributions to APV is mostly determined by the energy scale $Q$, i.e. 
\begin{eqnarray}
\label{eqn:APV:comparison}
\frac{ \Delta A^{\rm PV} ( ^{133}{\rm Cs} ) / A_{\rm SM}^{\rm PV} ( ^{133}{\rm Cs} ) } { \Delta A^{\rm PV} ( ^{12}{\rm C} ) / A_{\rm SM}^{\rm PV} ( ^{12}{\rm C} ) } (m_{Z'} \ll Q) & \ = \ &
\left( \frac{93\; {\rm MeV}}{30\; {\rm MeV}} \right)^2 ,
\end{eqnarray}
and the resulting APV limit is stronger than the P2 prospect in $e+ ^{12}$C scattering by a factor of
\begin{eqnarray}
\left( \frac{0.49\%}{0.77\%}  \right)^{1/2}
\left( \frac{93\; {\rm MeV}}{30\; {\rm MeV}}  \right) \ \simeq \ 2.47 \,.
\end{eqnarray}
In the limit of $m_{Z'} \gg Q$, comparison of the APV limits and the P2 prospects will be only dictated by the accuracies, i.e.\ the APV limit is weaker than the P2 sensitivity for $e + ^{12}$C by a factor of $\sqrt{0.49\% / 0.77\% }\simeq 0.80$. As shown in Fig.~\ref{fig:theta}, the APV constraints have excluded P2 sensitivities for $e+p$ scattering.
The limit from $\nu + e$ scattering  has  the same feature in the high $Z'$ mass range, and can go to masses beyond the ones from direct LHC searches, as shown in Fig.~\ref{fig:theta}. However, the P2 experiment can exceed the neutrino scattering limits. The resultant P2 ranges of $m_{Z'}$ and $\sin\theta$ are collected in Table~\ref{tab:limits2b}.

The P2 sensitivities for the most generic $U(1)'$ model with a kinetic mixing of $Z'$ with the SM $Z$ boson and with $g' Q' = 0$, as well as the corresponding existing limits, are presented in Fig.~\ref{fig:eps}. As shown in this figure, the P2 sensitivities for this case are not as competitive as others. This is because $Z'$ with a sizable kinetic mixing is more photon-like when $m_{Z'}$ decreases. In particular, based on the couplings in Tab.~\ref{tab:t}, in the limit of  $m_{Z'}\rightarrow 0$,
\begin{eqnarray}
g'_{e_L} - g'_{e_R} \ \propto \ \frac{m_{Z'}^2}{m_Z^2} \ \to \ 0 \,.
\end{eqnarray}
In fact, all the $g'_f$ couplings are proportional to the electric charge of $f$ in this limit, which implies that $Z'$ would not mediate parity-violating processes.  In this case, $\epsilon$ is mainly constrained by the modification of $Z$ couplings. As a result of the enhancement of $\Delta A^{\rm PV}$ due to the $(1-r_m)$ factor in the denominator of Eqs.~(\ref{eq:APV_eps}) and (\ref{eq:APV_epsC12}), there are dips at the $Z$ mass. However, the P2 sensitivities at the high $Z'$ mass range are still precluded by the electroweak precision data.

\begin{figure}[t]
\centering
\includegraphics[width=0.9\textwidth]{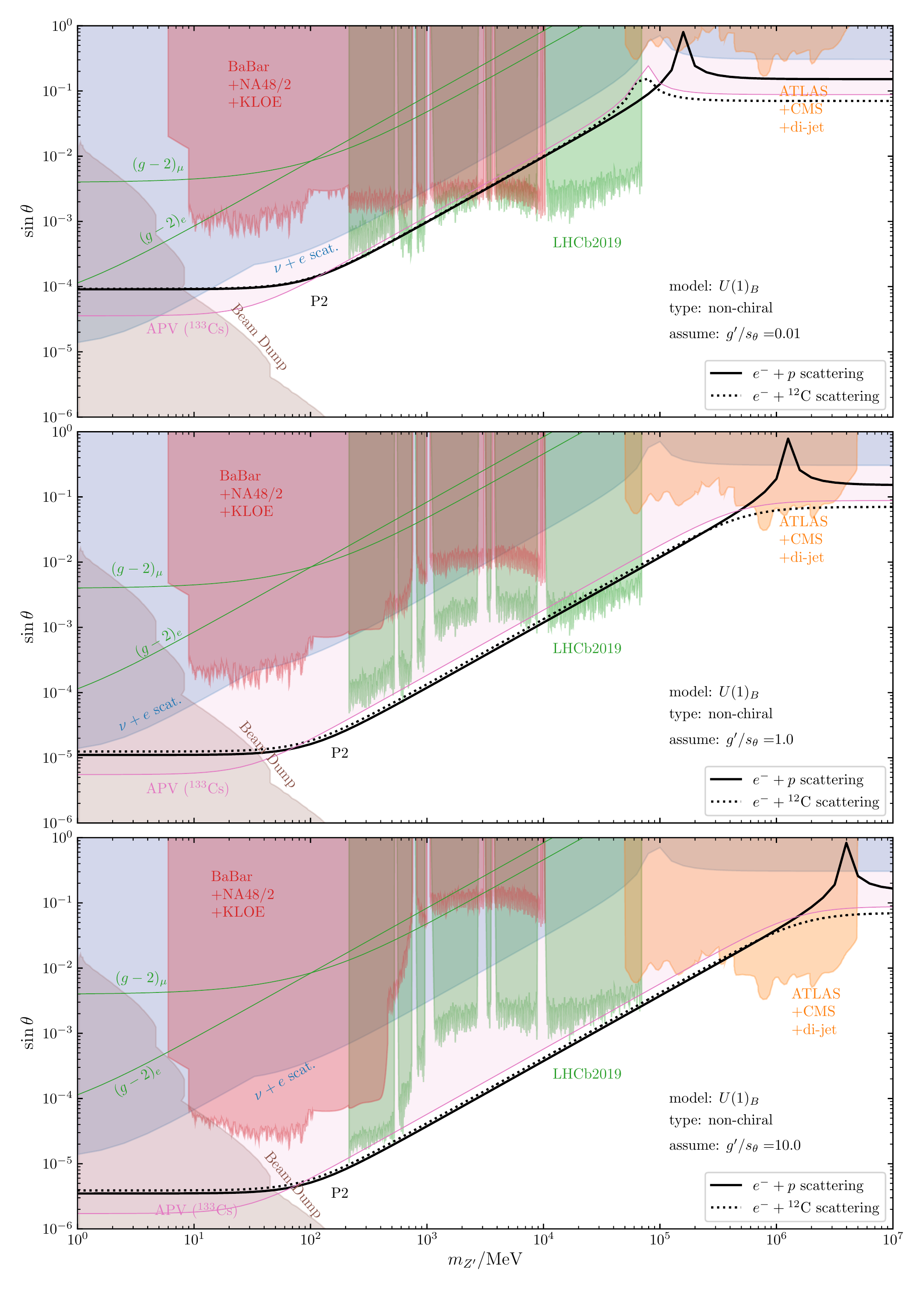}
\caption{Sensitivity of the P2 experiment to the chiral $U(1)_{B}$
model with mass mixing $\sin\theta$ and the three  values of $g'/\sin\theta = 0.01$ (upper), $1$ (middle) and $10$ (lower). The notations are the same as in Fig.~\ref{fig:gx}.
\label{fig:U1B}}
\end{figure}

The P2 sensitivities for the $U(1)_B$ model with $Z$-$Z'$ mass mixing and three  values of $g'/\sin\theta = 0.01$, $1$ and $10$ are shown respectively in the upper, middle and lower panels of Fig.~\ref{fig:U1B}. The master formula for $\Delta A^{\rm PV}$ is given in Eq.~(\ref{eq:n-1}). For the three benchmark values, the $R$ ratio is respectively 1.96, 196 and 1960 for  $e+p$ scattering, and $-0.34$, $-34$ and $-340$ for $e+^{12}$C scattering. Generally speaking,  when the $Z'$ mass is small such that $m_{Z}^2 \gtrsim m_{Z'}^2 + Q^2$, the P2 sensitivities of $\sin\theta$ can be significantly improved by a factor of $\sqrt{1+R}$. For sufficiently large $m_{Z'}$, i.e.
\begin{eqnarray}
m_{Z'} \ = \ \sqrt{1+R} \, m_{Z} \,,
\end{eqnarray}
the two terms in Eq.~(\ref{eq:n-1}) cancel each other for all three values of $g'$ in $e+p$ scattering, which turns out to happen respectively at
\begin{eqnarray}
m_{Z'} = 1.72 m_{Z} \;\; [157 \; {\rm GeV}]\,, \quad
14.1 m_{Z'} \;\; [1.28 \; {\rm TeV}]\,, \quad
44.3 m_{Z'} \;\; [4.03 \; {\rm TeV}]\,,
\end{eqnarray}
as indicated by the solid black peaks in Fig.~\ref{fig:U1B}. For the case of $g'/\sin\theta = 0.01$ in $e+^{12}$C scattering, the cancellation happens at
\begin{eqnarray}
m_{Z'} \ = \ 0.81 m_{Z} \;\; [74 \; {\rm GeV}]\,,
\end{eqnarray}
as shown by the peak of the dashed black line in the upper panel of Fig.~\ref{fig:U1B}. For the two cases of $g'/ \sin\theta =1$ and $10$ in $e + ^{12}$C scattering,  the factor $R$ is below $-1$ and there is no cancellation between the two terms in Eq.~\eqref{eq:n-1}. Therefore there are no peaks for the dashed curves in the middle and lower panels of Fig.~\ref{fig:U1B}.  In the limit of $m_{Z'} \gg \sqrt{1+R} \, m_Z$, the first term in the bracket of  Eq.~\eqref{eq:n-1} can be neglected, and the P2 sensitivity to $\sin\theta$ approaches a constant value, which is the same as in Fig.~\ref{fig:theta}. For the case of $g'/\sin\theta=0.01$, the current limits on the mass mixing angle $\sin\theta$ in the upper panel of Fig.~\ref{fig:U1B} are almost the same as in Fig.~\ref{fig:theta}, while the limits for the cases of $g/\sin\theta=1$ and 10 with larger gauge couplings tend to be more stringent, as presented in the middle and lower panels of Fig.~\ref{fig:U1B}. In light of all the limits, the ranges of $m_{Z'}$ and $\sin\theta$ that can be probed at P2 for the $U(1)_B$ model are collected in Table~\ref{tab:limits2b}.

\begin{figure}[t]
\centering
\includegraphics[width=0.9\textwidth]{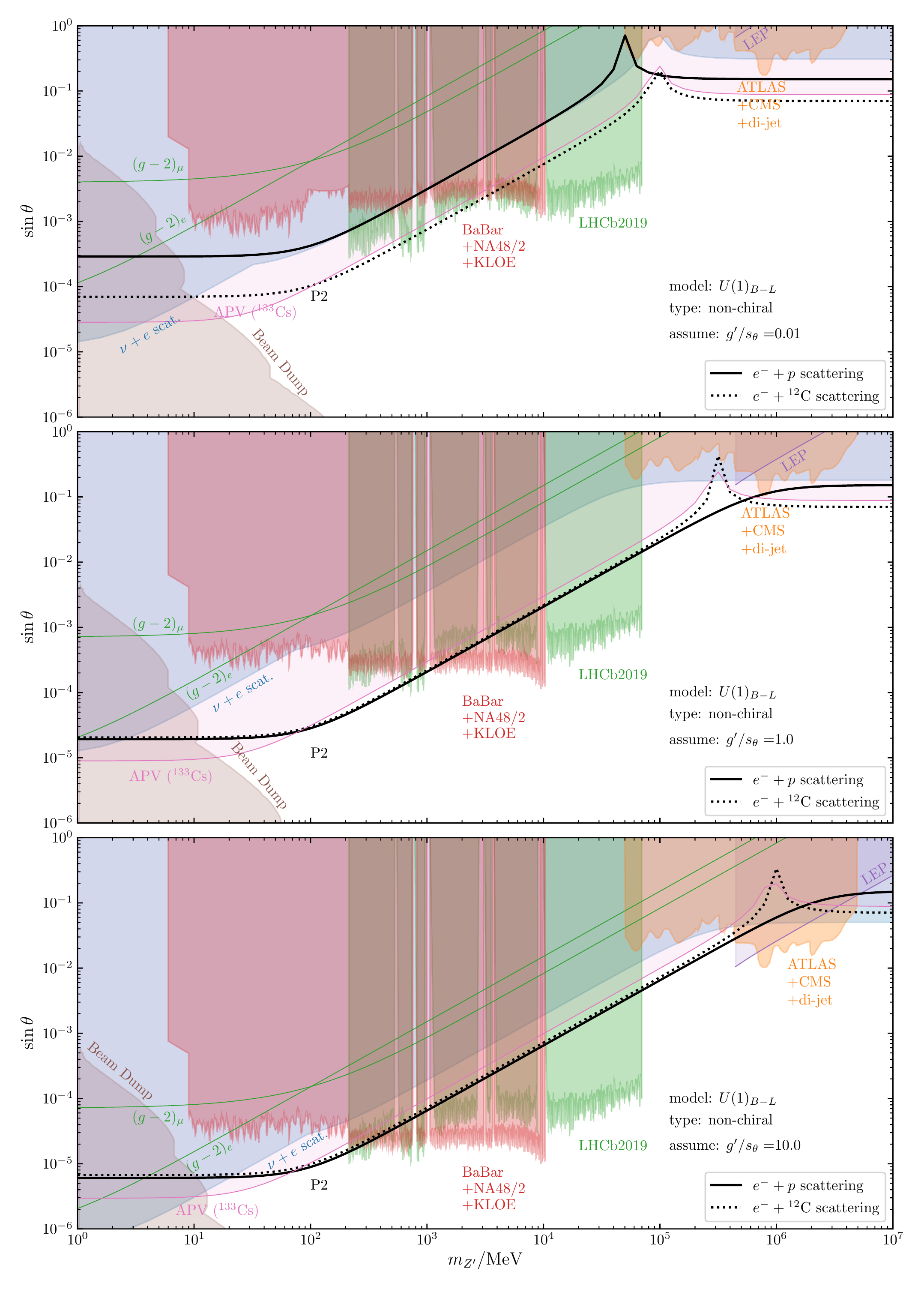}
\caption{Similar to Fig.~\ref{fig:U1B} but for the  $U(1)_{B-L}$ model.  \label{fig:U1B-L}}
\end{figure}

The $U(1)_{B-L}$ model with $Z$-$Z'$ mass mixing is quite similar to the $U(1)_B$ model, and the corresponding P2 sensitivities and current limits are presented in Fig.~\ref{fig:U1B-L}, with the three benchmark values of $g'/\sin\theta = 0.01$, $1$ and $10$ respectively in the upper, middle and lower panels. The key difference between the $U(1)_B$ and $U(1)_{B-L}$ models is the $R$ factor relation in Eq.~(\ref{eqn:Rfactor}). As a result, for $e+p$ scattering in the $U(1)_{B-L}$ model, we  have a disappearing P2 sensitivity only at
\begin{eqnarray}
m_{Z'} \ = \ 0.57 m_{Z} \;\; [51 \; {\rm GeV}]
\end{eqnarray}
for $g'/\sin\theta = 0.01$, while for larger $g' = 1$ and $10$ the factor $R$ is below $-1$ and there is no peak. Regarding $e+ ^{12}$C scattering,  P2 looses sensitivity at the following $Z'$ masses:
\begin{eqnarray}
m_{Z'} \ = \ 1.06 m_{Z} \;\; [96 \; {\rm GeV}]\,, \quad
3.57 m_{Z'} \;\; [325 \; {\rm GeV}]\,, \quad
10.9 m_{Z'} \;\; [990 \; {\rm GeV}]
\end{eqnarray}
for the three values of $g'/\sin\theta = 0.01$, $1$ and $10$, respectively. The P2 sensitivities of $m_{Z'}$ and $\sin\theta$ are also summarized in Table~\ref{tab:limits2b}.

For the $U(1)_B$ and $U(1)_{B-L}$ models, the couplings of the $Z'$ boson to proton and $^{12}$C receive  contributions from the direct coupling and the $Z$-$Z'$ mixing, which are proportional to $g'$ and $\sin\theta$, respectively. When the gauge coupling $g'$ is sufficiently large, the couplings of $Z'$ to proton and $^{12}$C will be dominated by the direct coupling $g'$, which is different from the pure $Z$-$Z'$ mass mixing case in Fig.~\ref{fig:theta}. On the other hand, for sufficiently large $g'$ and light $Z'$, the new physics contribution to the asymmetry will be proportional to the $R$ factor, i.e.\  $\Delta A^{\rm PV} \propto R$ in Eq.~(\ref{eq:n-1}). Taking into account the different factors for proton and $^{12}$C in Eq.~(\ref{eq:factors}), for the cases of $g'/\sin\theta = 0.01$, 1 and 10 for $U(1)_B$ and the cases of $g'/\sin\theta = 1$ and 10 for $U(1)_{B-L}$, the P2 sensitivities for low $m_{Z'}$ are {\it accidentally} roughly the same for proton and $^{12}$C, as shown in Figs.~\ref{fig:U1B} and \ref{fig:U1B-L}.

\section{Conclusion}
\label{sec:concl}

With longitudinally polarized electron scattering off proton and $^{12}$C, the  P2 experiment can perform a measurement of the parity-violating asymmetry $A^{\rm PV}$ at the unprecedented precision of below per cent level, which would result in a precision of $3.3 \times 10^{-4}$ for the weak mixing angle $s_W$ at sub-GeV scales. Such high-precision measurements have excellent sensitivities to potential BSM physics, as we have illustrated in this paper, taking as a case study arguably the most straightforward  scenario: new $Z'$ bosons. We assumed chiral and non-chiral $U(1)'$ models, for the latter case assuming  kinetic or mass $Z$-$Z'$ mixing. For the chiral case, where the $Z'$ couplings to left- and right-handed fermions are different, we adopt three $U(1)'$ models, i.e.\ the $U(1)'_L$, $U(1)'_R$ and $U(1)'_X$ models given in Table~\ref{tab:chiral_models}. For the non-chiral models, we first consider a generic $U(1)'$ model with either mass ($\sin\theta$) or kinetic ($\epsilon$) mixing of $Z'$ with the $Z$ boson and  setting $g'$ to zero, and then generalize to the $U(1)_B$ and $U(1)_{B-L}$ models with three different benchmark values of $g'/\sin\theta = 0.01$, $1$ and $10$.

For all  chiral and non-chiral models, the leading-order new physics contributions are dominated by the interference  terms of the BSM diagrams with the SM terms, i.e.\  terms proportional to the gauge couplings $g^{\prime 2}$, the mass mixing angle $\sin^2\theta$ or the kinetic mixing parameter $\epsilon^2$. The P2 sensitivities for these $U(1)'$ models, as well as the current limits, are presented in Figs.~\ref{fig:gx} to \ref{fig:U1B-L}. Let us summarize our main results from these figures:
\begin{itemize}
    \item It is a general feature that when the $Z'$ mass is smaller than the energy scale $Q \simeq 93$ MeV of the P2 experiment, i.e.\ $m_{Z'} \lesssim Q$, the P2 sensitivities will be independent of the $Z'$ mass, as shown in the second and third columns of Table~\ref{tab:limits}. However, the P2 prospects in the low $Z'$ mass range are mostly precluded by the beam-dump experiments, and some are also limited by APV measurements, neutrino scattering data, electron and muon $g-2$, and electroweak precision data.

    \item In the large $Z'$ mass limit, for the chiral $U(1)'$ models and the kinetic mixing case of non-chiral models, the P2 sensitivities are proportional to $g'/m_{Z'}$ or $\epsilon/m_{Z'}$. In other words, the P2 experiment can probe effectively a UV cutoff scale $\Lambda = m_{Z'} / g'$ or $\Lambda = m_{Z'} / \epsilon$. For the chiral models, the P2 experiment has sensitivities to a $Z'$ mass up to 79 TeV in the proton mode and up to 90 TeV in $^{12}$C mode, assuming the perturbative gauge coupling limit of $g' = 4\pi$, as shown in Table~\ref{tab:limits2a}. For the case of mass mixing $\sin\theta$ in non-chiral models, when the $Z'$ boson is very heavy, say $m_{Z'} \gtrsim \sqrt{1+R} \, m_{Z}$, the first term in the bracket of Eq.~(\ref{eq:m-5}) or (\ref{eq:n-1}) will be highly suppressed, and the P2 prospects for $\sin\theta$ will be independent of the $Z'$ mass, which is very different from the cases above. The P2 sensitivities in the limit of heavy $Z'$ bosons are collected in the fourth and fifth columns of Table~\ref{tab:limits}.

    One can see that for most of the cases in this paper, i.e.\ the three chiral models, generic $U(1)'$ with mass mixing ($\sin\theta$), the $U(1)_B$ model, and the $U(1)_{B-L}$ model with  $g'/\sin\theta = 0.01$ and $1$, the P2 experiment can probe a high UV scale $\Lambda$ or a mass mixing angle $\sin\theta$ that is currently not constrained and even goes beyond the direct search limits from LHC, in particular for $e+^{12}$C scattering for most of the models. It is promising that a superheavy $Z'$ boson can be directly searched for at future 100 TeV colliders~\cite{Arkani-Hamed:2015vfh, Golling:2016gvc}, which is largely complementary to the prospects at the high-precision P2 experiment. The P2 prospects for a heavy $Z'$ boson in the kinetic mixing $U(1)'$ model is precluded by electroweak precision data, while the $U(1)_{B-L}$ model with a large gauge coupling $g'/\sin\theta$ is excluded by  neutrino scattering data for a heavy $Z'$ boson.

    \item One may wonder whether the P2 experiment running with a $^{12}{\rm C}$ target could have significantly improved sensitivity to new physics compared to that with a proton target. From Figs.~\ref{fig:gx} to \ref{fig:U1B-L}, we can see that this is rather model-dependent, subject to the quantum numbers of SM fermions under the $U(1)'$ gauge group for the chiral $U(1)'$ model, and depending on the gauge coupling $g'$ for the non-chiral models. For the $U(1)'_R$ model, there is  no sensitivity with the $^{12}$C target, which is due to an accidental cancellation of the quantum numbers in Eq.~(\ref{eq:m-11}). For the $U(1)'_L$ model, the $^{12}$C target can improve the sensitivity from the proton target by a factor of 1.4, whereas for the $U(1)'_X$ model the prospects at proton and $^{12}$C targets are roughly the same.
    For non-chiral models, the $^{12}{\rm C}$ target does exhibit better sensitivity when $g'$ is small; see Figs.~\ref{fig:theta}, \ref{fig:U1B} and \ref{fig:U1B-L}. In  the limit of $g'=0$, as shown in Figs.~\ref{fig:theta}, this improvement can reach a factor of 2.\newline
    In any case, the comparison of electron-proton and electron-$^{12}$C scatterings can obviously help in distinguishing models. Comparing further with the parity asymmetry in polarized electron-electron (M{\o}ller) scattering provides further handle on identifying the underlying physics~\cite{Benesch:2014bas}.

    \item The nuclear  6$S_{1/2}$ $-$ 7$S_{1/2}$ transition in $^{133}$Cs provides stringent limits on the chiral and non-chiral models which are comparable to the P2 prospects. Depending on model details, the P2 experiment can achieve better sensitivities than the APV measurements, in particular if the $Z'$ boson is relatively heavy.

\end{itemize}

Taking into account all relevant existing constraints,  we find that P2 can probe a broad range of $m_{Z'}$ in the $U(1)'$ models considered in this paper, which are collected in Table~\ref{tab:limits2a} for chiral models and Table~\ref{tab:limits2b} for non-chiral models.

One should note that the specific $U(1)'$ models considered here are only for illustration purposes, and the analysis in this paper can be easily generalized to other $U(1)'$ models, by taking  different choices of the quantum numbers $x$ and $y$ in Eq.~(\ref{eq:m-8}), or to flavor-sensitive scenarios such as the $L_\mu - L_\tau$ model.


\begin{acknowledgments}
We thank Niklaus Berger, Jeff Dror, Bhaskar Dutta and Krishna Kumar for helpful discussions.
The work of B.D.\ is supported in part by the US Department of Energy under Grant No. DE-SC0017987, by the Neutrino Theory Network Program, and by a Fermilab Intensity Frontier Fellowship. X.J.X.\ is supported by the ``Probing dark matter with neutrinos'' ULB-ARC convention and by the F.R.S./FNRS under the Excellence of Science (EoS) project No.\ 30820817 - be.h ``The $H$ boson gateway to physics beyond the Standard Model''. Y.Z. is partially supported by ``the Fundamental Research Funds for the Central Universities''.
\end{acknowledgments}

\appendix

\section{The contribution of the axial-vector coupling}
\label{sec:gA}

In this appendix we show that the effect of $g_{A}$ (and $g'_{A}$)
is suppressed by a factor of $E_{e}/m_{N}$ (which in P2 is about 0.16 for
electron-proton scattering, and 0.013 for electron-$^{12}$C scattering)
compared to the leading-order contribution to $A^{{\rm PV}}$ {from $g_V$ (and $g_V'$)}.
To evaluate the contributions
of $g_{A}$ and $g'_{A}$, we first write down the full amplitude:
\begin{eqnarray}
i{\cal M}_{L,R} & \ \propto \ & [\overline{u_{3}}\gamma^{\mu}P_{L,R}u_{1}][\overline{u_{4}}\gamma_{\mu}u_{2}]\frac{-e^{2}}{q^{2}}\nonumber \\
 & & +  [\overline{u_{3}}\gamma^{\mu}P_{L,R}g_{L,R}u_{1}][\overline{u_{4}}\gamma_{\mu}(g_{V}+g_{A}\gamma^{5})u_{2}]\frac{1}{q^{2}-m_{Z}^{2}}\nonumber \\
 & & +  [\overline{u_{3}}\gamma^{\mu}P_{L,R}g'_{L,R}u_{1}][\overline{u_{4}}\gamma_{\mu}(g'_{V}+g'_{A}\gamma^{5})u_{2}]\frac{1}{q^{2}-m_{Z'}^{2}}\, .\label{eq:x-4-1}
\end{eqnarray}
Hence Eq.~(\ref{eq:x-7}) is modified to
\begin{eqnarray}
i{\cal M}_{L,R} & \ \propto \ & [\overline{u_{3}}\gamma^{\mu}P_{L,R}u_{1}]\left[\overline{u_{4}}\gamma_{\mu}(G_{L,R}+G_{A}\gamma^{5})u_{2}\right]\, ,\label{eq:x-7-1}
\end{eqnarray}
where
\begin{equation}
G_{A} \ = \ \frac{g_{L,R}g_{A}}{q^{2}-m_{Z}^{2}}+\frac{g'_{L,R}g'_{A}}{q^{2}-m_{Z'}^{2}} \, .\label{eq:x-6-1}
\end{equation}
Applying the standard trace technology, we obtain:
\begin{eqnarray}
|{\cal M}_{L,R}|^{2} & \ \propto \ & G_{L,R}^{2}\left(2-2y-4r_{2}y-r_{1}^{2}+4r_{2}^{2}y^{2}\right)
-8G_{L,R}G_{A}\left(1-r_{2}y\right)r_{2}y\nonumber \\
 &  & +G_{A}^{2}\left(2+2y-4r_{2}y+r_{1}^{2}+4r_{2}^{2}y^{2}\right),\label{eq:x-8}
\end{eqnarray}
where $r_{1}\equiv m_{e}/E_{e}\approx 3\times10^{-3}$,
$r_{2}\equiv E_{e}/m_{N}\approx 0.16$ (for $e+p$ scattering) or 0.013 (for $e+{}^{12}\text{C}$ scattering),
and $y\equiv -q^{2}/4E_{e}^{2}\approx 0.09$.
The second term in Eq.~(\ref{eq:x-8}) proportional to $G_A$ implies that the axial-vector contribution
is suppressed by $r_2y$.
The third term   proportional to $G_{A}^{2}$, though
not  suppressed, cancels in $|{\cal M}_{R}|^{2}-|{\cal M}_{L}|^{2}$.


\bibliographystyle{JHEP}
\bibliography{ref}

\end{document}